\documentclass[aip,jcp,reprint,noshowkeys,superscriptaddress]{revtex4-1}
\usepackage{graphicx,dcolumn,bm,xcolor,microtype,multirow,amsmath,amssymb,amsfonts,physics,mhchem,xspace,subfigure}

\usepackage[utf8]{inputenc}
\usepackage[T1]{fontenc}
\usepackage{txfonts}

\usepackage[
	colorlinks=true,
    citecolor=blue,
    breaklinks=true
	]{hyperref}
\definecolor{darkgreen}{HTML}{009900}
\usepackage[normalem]{ulem}

\newcommand{\ie}{\textit{i.e.}}

\newcommand{\fnm}{\footnotemark}
\newcommand{\fnt}{\footnotetext}
\newcommand{\tabc}[1]{\multicolumn{1}{c}{#1}}

\newcommand{\ai}[1]{\hat{a}_{#1}}
\newcommand{\aic}[1]{\hat{a}^{\dagger}_{#1}}

\newcommand{\CBS}{\text{CBS}}

\newcommand{\efci}[0]{E_{\text{FCI}}^{\Bas}}

\newcommand{\efuncbasisFCI}[0]{\bar{E}^\Bas[\denFCI]}

\newcommand{\efuncbasis}[0]{\bar{E}^\Bas[\den]}
\newcommand{\efuncden}[1]{\bar{E}^\Bas[#1]}
\newcommand{\efuncdenpbe}[1]{\bar{E}_{\text{}}^\Bas[#1]}
\newcommand{\efuncdenpbeAB}[1]{\bar{E}_{\text{A}+\text{B}}^\Bas[#1]}

\newcommand{\ecmd}[0]{\bar{\varepsilon}_{\text{c,md}}^{\text{sr},\text{PBE}}}
\newcommand{\psibasis}[0]{\Psi^{\basis}}
\newcommand{\BasFC}{\mathcal{V}}  

\newcommand{\pbeuegXi}{\text{PBE-UEG}}

\newcommand{\argrpbeuegXi}[0]{\den(\br{}),\tilde{\zeta}(\br{}),s(\br{}),\ntwo^{\text{UEG}}(\br{}),\mu_{\text{}}(\br{})}
\newcommand{\pbeontXi}{\text{PBE-OT}}

\newcommand{\argrpbeontXi}[0]{\den(\br{}),\tilde{\zeta}(\br{}),s(\br{}),\ntwoextrapcas(\br{}),\mu_{\text{}}^{}(\br{})}
\newcommand{\pbeontns}{\text{SU-PBE-OT}}

\newcommand{\argrpbeontns}[0]{\den(\br{}),0,s(\br{}),\ntwoextrapcas(\br{}),\mu_{\text{}}^{}(\br{})}

\newcommand{\argepbe}[0]{\den,\zeta,s}
\newcommand{\argebasis}[0]{\den,\zeta,\ntwo,\mu}
\newcommand{\argecmd}[0]{\den,\zeta,s,\ntwo,\mu}

\newcommand{\argrebasis}[0]{\denr,\zeta(\br{}),s(\br{}),\ntwo(\br{}),\mu(\br{})}

\newcommand{\bfr}[1]{{\bf r}_{#1}}
\newcommand{\dr}[1]{\text{d}\bfr{#1}}
\newcommand{\rr}[2]{\bfr{#1}, \bfr{#2}}

\newcommand{\murpsi}[0]{\mu_{\wf{}{\Bas}}({\bf r})}
\newcommand{\murcas}[0]{\mu_{\text{CASSCF}}({\bf r})}
\newcommand{\mucas}[0]{\mu_{\text{CASSCF}}}
\newcommand{\murcipsi}[0]{\mu_{\text{CIPSI}}({\bf r})}
\newcommand{\mucipsi}[0]{\mu_{\text{CIPSI}}}

\newcommand{\ntwo}[0]{n_{2}}

\newcommand{\ntwoextrap}[0]{\mathring{n}_{2}}
\newcommand{\ntwoextrapcas}[0]{\mathring{n}_2^{\text{}}}
\newcommand{\mur}[0]{\mu({\bf r})}

\newcommand{\murpsival}[0]{\mu_{\wf{}{\Bas}}^{\text{val}}({\bf r})}

\newcommand{\wbasis}[0]{W_{\wf{}{\Bas}}(\bfr{1},\bfr{2})}
\newcommand{\wbasiscoal}[0]{W_{\wf{}{\Bas}}(\bfr{},\bfr{})}
\newcommand{\wbasisval}[0]{W_{\wf{}{\Bas}}^{\text{val}}(\bfr{1},\bfr{2})}
\newcommand{\fbasis}[0]{f_{\wf{}{\Bas}}(\bfr{1},\bfr{2})}
\newcommand{\fbasisval}[0]{f_{\wf{}{\Bas}}^{\text{val}}(\bfr{1},\bfr{2})}

 \newcommand{\twodmrdiagpsi}[0]{ n_{2,{\wf{}{\Bas}}}(\rr{1}{2})}
 \newcommand{\twodmrdiagpsival}[0]{n_{2,\wf{}{\Bas}}^{\text{val}}(\rr{1}{2})}

\newcommand{\wbasiscoalval}{W_{\wf{}{\Bas}}^{\text{val}}({\bf r},{\bf r})}

\newcommand{\denFCI}[0]{\den^{\Bas}_{\text{FCI}}}

\newcommand{\den}[0]{{n}}

\newcommand{\denr}[0]{{n}({\bf r})}

\newcommand{\psifci}[0]{\Psi^{\Bas}_{\text{FCI}}}

\newcommand{\kinop}[0]{\hat{T}}

\newcommand{\weeop}[0]{\hat{W}_{\text{ee}}}

\newcommand{\lr}{\text{lr}}

\newcommand{\V}[2]{V_{#1}^{#2}}

\newcommand{\n}[2]{n_{#1}^{#2}}

\newcommand{\wf}[2]{\Psi_{#1}^{#2}}

\newcommand{\SO}[2]{\phi_{#1}(\br{#2})}

\newcommand{\Bas}{\mathcal{B}}
\newcommand{\basis}{\mathcal{B}}

\newcommand{\Cor}{\mathcal{C}}

\renewcommand{\d}{\text{d}}

\newcommand{\Gam}[2]{\Gamma_{#1}^{#2}}
\newcommand{\isEquivTo}[1]{\underset{#1}{\sim}}

\newcommand{\br}[1]{{\mathbf{r}_{#1}}}

\newcommand{\ontopcas}{\langle n_{2,\text{CASSCF}} \rangle}
\newcommand{\ontopextrap}{\langle \mathring{n}_{2,\text{CASSCF}} \rangle}
\newcommand{\ontopextrapcipsi}{\langle \mathring{n}_{2,\text{CIPSI}} \rangle}
\newcommand{\ontopcipsi}{\langle n_{2,\text{CIPSI}} \rangle}
\newcommand{\muaverage}{\langle \mucas  \rangle}
\newcommand{\muaveragecipsi}{\langle \mucipsi  \rangle}

\newcommand{\LCT}{Laboratoire de Chimie Th\'eorique (UMR 7616), Sorbonne Universit\'e, CNRS, Paris, France}

\newcommand{\LCPQ}{Laboratoire de Chimie et Physique Quantiques (UMR 5626), Universit\'e de Toulouse, CNRS, UPS, France}
\newcommand{\IUF}{Institut Universitaire de France, Paris, France}

\begin{document}	

\title{A basis-set error correction based on density-functional theory for strongly correlated molecular systems}

\author{Emmanuel Giner}
\email{emmanuel.giner@lct.jussieu.fr}
\affiliation{\LCT}
\author{Anthony Scemama}
\affiliation{\LCPQ}
\author{Pierre-Fran\c{c}ois Loos}
\email{loos@irsamc.ups-tlse.fr}
\affiliation{\LCPQ}
\author{Julien Toulouse}
\email{toulouse@lct.jussieu.fr}
\affiliation{\LCT}
\affiliation{\IUF}

\date{April 7, 2020}

\begin{abstract}
We extend to strongly correlated molecular systems the recently introduced basis-set incompleteness correction based on density-functional theory (DFT) [E. Giner \textit{et al.}, \href{https://doi.org/10.1063/1.5052714}{J. Chem. Phys. \textbf{149}, 194301 (2018)}]. This basis-set correction relies on a mapping between wave-function calculations in a finite basis set and range-separated DFT (RSDFT) through the definition of an effective non-divergent interaction corresponding to the electron-electron Coulomb interaction projected in the finite basis set. This enables the use of RSDFT-type complementary density functionals to recover the dominant part of the short-range correlation effects missing in this finite basis set. To study both weak and strong correlation regimes we consider the potential energy curves of the \ce{H10}, \ce{N2}, \ce{O2}, and \ce{F2} molecules up to the dissociation limit, and we explore various approximations of complementary functionals fulfilling two key properties: spin-multiplet degeneracy (\ie, independence of the energy with respect to the spin projection $S_z$) and size consistency. Specifically, we investigate the dependence of the functional on different types of on-top pair densities and spin polarizations. The key result of this study is that the explicit dependence on the on-top pair density allows one to completely remove the dependence on any form of spin polarization without any significant loss of accuracy. Quantitatively, we show that the basis-set correction reaches chemical accuracy on atomization energies with triple-$\zeta$ quality basis sets for most of the systems studied here. Also, the present basis-set incompleteness correction provides smooth potential energy curves along the whole range of internuclear distances.
\end{abstract}

\maketitle

%%%%%%%%%%%%%%%%%%%%%%%%
\section{Introduction}
%%%%%%%%%%%%%%%%%%%%%%%%
The general goal of quantum chemistry is to provide reliable theoretical tools to explore the rich area of chemistry. More specifically, developments in quantum chemistry primarily aim at accurately computing the electronic structure of molecular systems. Despite intense developments, no definitive solution to this problem has been found. The theoretical challenge to tackle belongs to the quantum many-body problem, due to the intrinsic quantum nature of the electrons and the Coulomb repulsion between them. This so-called electronic correlation problem corresponds to finding a solution to the Schr\"odinger equation for a $N$-electron system, and two main roads have emerged to approximate this solution: wave-function theory (WFT) \cite{Pop-RMP-99} and density-functional theory (DFT). \cite{Koh-RMP-99} Although both WFT and DFT spring from the same Schr\"odinger equation, they rely on very different formalisms, as the former deals with the complicated $N$-electron wave function whereas the latter focuses on the much simpler one-electron density. In its Kohn-Sham (KS) formulation, \cite{KohSha-PR-65} the computational cost of DFT is very appealing since it is a simple mean-field procedure. Therefore, although continued efforts have been made to reduce the computational cost of WFT, DFT still remains the workhorse of quantum computational chemistry. 

The difficulty of obtaining a reliable theoretical description of a given chemical system can be roughly categorized by the strength of the electronic correlation. The so-called weakly correlated systems, such as closed-shell organic molecules near their equilibrium geometry, are typically dominated by correlation effects which do not affect the qualitative mean-field picture of the system. These weak-correlation effects can be either short range (near the electron-electron coalescence points) \cite{HatKloKohTew-CR-12} or long range (London dispersion interactions). \cite{AngDobJanGou-BOOK-20} The theoretical description of weakly correlated systems is one of the most concrete achievement of quantum chemistry, and the main remaining challenge for these systems is to push the limit of the chemical system size that can be treated.  The case of the so-called strongly correlated systems, which are ubiquitous in chemistry, is more problematic as they exhibit a much more complex electronic structure. For example, transition metal complexes, low-spin open-shell systems, covalent bond breaking situations have all in common that they cannot be even qualitatively described by a single electronic configuration. It is now clear that the usual semilocal density-functional approximations of KS DFT fail to accurately describe these situations \cite{GorSeiSav-PCCP-08,GagTruLiCarHoyBa-ACR-17} and WFT is king for the treatment of strongly correlated systems. 

In practice, WFT uses a finite one-electron basis set. The exact solution of the Schr\"odinger equation within this basis set is then provided by full configuration interaction (FCI) which consists in a linear-algebra eigenvalue problem with a dimension scaling exponentially with the system size. Due to this exponential growth of the FCI computational cost, introducing approximations is necessary, with at least two difficulties for strongly correlated systems: i) the qualitative description of the wave function is determined by a primary set of electronic configurations (whose size can scale exponentially in many cases) among which near degeneracies and/or strong interactions appear in the Hamiltonian matrix; ii) the quantitative description of the system requires also to account for weak correlation effects which involve many other electronic configurations with typically much smaller weights in the wave function. Addressing simultaneously these two issues is a rather complicated task for a given approximate WFT method, especially if one adds the requirement of satisfying formal properties, such as spin-multiplet degeneracy (\ie, independence of the energy with respect to the spin projection $S_z$) and size consistency.

Beside the difficulties of accurately describing the molecular electronic structure within a given basis set, a crucial limitation of WFT methods is the slow convergence of the energy (and related properties) with respect to the size of the one-electron basis set. As initially shown by the seminal work of Hylleraas \cite{Hyl-ZP-29} and further developed by Kutzelnigg and coworkers, \cite{Kut-TCA-85,KutKlo-JCP-91, NogKut-JCP-94} the main convergence problem originates from the divergence of the electron-electron Coulomb interaction at the coalescence point, which induces a discontinuity in the first derivative of the exact wave function (the so-called electron-electron cusp). Describing such a discontinuity with an incomplete one-electron basis set is impossible and, as a consequence, the convergence of the computed energies and properties are strongly affected. To alleviate this problem, extrapolation techniques have been developed, either based on a partial-wave expansion analysis, \cite{HelKloKocNog-JCP-97,HalHelJorKloKocOlsWil-CPL-98} or more recently based on perturbative arguments. \cite{IrmHulGru-PRL-19,IrmGru-JCP-2019} A more rigorous approach to tackle the basis-set convergence problem is provided by the so-called explicitly correlated F12 (or R12) methods \cite{Ten-TCA-12,TenNog-WIREs-12,HatKloKohTew-CR-12, KonBisVal-CR-12, GruHirOhnTen-JCP-17, MaWer-WIREs-18} which introduce a geminal function depending explicitly on the interelectronic distance. This ensures a correct representation of the Coulomb correlation hole around the electron-electron coalescence point, and leads to a much faster convergence of the energy than usual WFT methods. For instance, using the explicitly correlated version of coupled cluster with singles, doubles, and perturbative triples [CCSD(T)] in a triple-$\zeta$ basis set is equivalent to using a quintuple-$\zeta$ basis set with the usual CCSD(T) method, \cite{TewKloNeiHat-PCCP-07} although a computational overhead is introduced by the auxiliary basis set needed to compute the three-electron integrals involved in F12 theory. \cite{BarLoo-JCP-17} In addition to the computational cost, a possible drawback of F12 theory is its rather complex formalism which requires non-trivial developments for adapting it to a new method. For strongly correlated systems, several multi-reference methods have been extended to explicit correlation (see, for example, Refs.~\onlinecite{Ten-CPL-07,ShiWer-JCP-10,TorKniWer-JCP-11,DemStanMatTenPitNog-PCCP-12,GuoSivValNee-JCP-17}), including approaches based on the so-called universal F12 theory which are potentially applicable to any electronic-structure computational methods. \cite{TorVal-JCP-09,KonVal-JCP-11,HauMaoMukKlo-CPL-12,BooCleAlaTew-JCP-12}

An alternative way to improve the convergence towards the complete-basis-set (CBS) limit is to treat the short-range correlation effects within DFT and to use WFT methods to deal only with the long-range and/or strong correlation effects. A rigorous approach achieving this mixing of DFT and WFT is range-separated DFT (RSDFT) (see Ref.~\onlinecite{TouColSav-PRA-04} and references therein) which relies on a decomposition of the electron-electron Coulomb interaction in terms of the interelectronic distance thanks to a range-separation parameter $\mu$. The advantage of this approach is at least two-fold: i) the DFT part deals primarily with the short-range part of the Coulomb interaction, and consequently the usual semilocal density-functional approximations are more accurate than for standard KS DFT; ii) the WFT part deals only with a smooth non-divergent interaction, and consequently the wave function has no electron-electron cusp \cite{GorSav-PRA-06} and the basis-set convergence is much faster. \cite{FraMusLupTou-JCP-15} A number of approximate RSDFT schemes have been developed involving single-reference \cite{AngGerSavTou-PRA-05, GolWerSto-PCCP-05, TouGerJanSavAng-PRL-09,JanHenScu-JCP-09, TouZhuSavJanAng-JCP-11, MusReiAngTou-JCP-15,KalTou-JCP-18,KalMusTou-JCP-19} and multi-reference \cite{LeiStoWerSav-CPL-97, FroTouJen-JCP-07, FroCimJen-PRA-10, HedKneKieJenRei-JCP-15, HedTouJen-JCP-18, FerGinTou-JCP-18} WFT methods. Nevertheless, there are still some open issues in RSDFT, such as remaining fractional-charge and fractional-spin errors in the short-range density functionals \cite{MusTou-MP-17} or the dependence of the quality of the results on the value of the range-separation parameter $\mu$.

Building on the development of RSDFT, a possible solution to the basis-set convergence problem has been recently proposed by some of the present authors~\cite{GinPraFerAssSavTou-JCP-18} in which RSDFT functionals are used to recover only the correlation effects outside a given basis set. The key point here is to realize that a wave function developed in an incomplete basis set is cuspless and could also originate from a Hamiltonian with a non-divergent long-range electron-electron interaction. Therefore, a mapping with RSDFT can be performed through the introduction of an effective non-divergent interaction representing the usual electron-electron Coulomb interaction projected in an incomplete basis set. First applications to weakly correlated molecular systems have been successfully carried out, \cite{LooPraSceTouGin-JCPL-19} together with extensions of this approach to the calculations of excitation energies \cite{GinSceTouLoo-JCP-19} and ionization potentials. \cite{LooPraSceGinTou-JCTC-20} The goal of the present work is to further develop this approach for the description of strongly correlated systems. 

The paper is organized as follows. In Sec.~\ref{sec:theory}, we recall the mathematical framework of the basis-set correction and we present its extension for strongly correlated systems. In particular, our focus is primarily set on imposing two key formal properties which are highly desirable in the context of strong correlation: spin-multiplet degeneracy and size consistency. To do this, we introduce i) new functionals using different flavors of spin polarizations and on-top pair densities, and ii) an effective electron-electron interaction based on a multiconfigurational wave function. This generalizes the method used in previous works on weakly correlated systems\cite{GinPraFerAssSavTou-JCP-18,LooPraSceTouGin-JCPL-19} for which it was sufficient to use an effective interaction based on a single-determinant wave function and a functional depending only on the usual density, reduced density gradient and spin polarization. 
Then, in Sec.~\ref{sec:results}, we apply the method to the calculation of the potential energy curves of the \ce{H10}, \ce{N2}, \ce{O2}, and \ce{F2} molecules up to the dissociation limit. Finally, we conclude in Sec.~\ref{sec:conclusion}. 

%%%%%%%%%%%%%%%%%%%%%%%%
\section{Theory}
\label{sec:theory}
%%%%%%%%%%%%%%%%%%%%%%%%
As the theory behind the present basis-set correction has been exposed in details in Ref.~\onlinecite{GinPraFerAssSavTou-JCP-18}, we only briefly recall the main equations and concepts needed for this study in Secs.~\ref{sec:basic}, \ref{sec:wee}, and \ref{sec:mur}. More specifically, in Sec.~\ref{sec:basic}, we recall the basic mathematical framework of the present theory by introducing the complementary functional to a basis set. Section \ref{sec:wee} introduces the effective non-divergent interaction in the basis set, which leads us to the definition of the effective \textit{local} range-separation function in Sec.~\ref{sec:mur}. Then, Sec.~\ref{sec:functional} exposes the new approximate RSDFT-based complementary correlation functionals. The generic form of such functionals is exposed in Sec.~\ref{sec:functional_form}, their properties in the context of the basis-set correction are discussed in Sec.~\ref{sec:functional_prop}, and the specific requirements for strong correlation are discussed in Sec.~\ref{sec:requirements}. Finally, the actual functionals used in this work are introduced in Sec.~\ref{sec:def_func}. 

\subsection{Basic theory}
\label{sec:basic}

The exact ground-state energy $E_0$ of a $N$-electron system can, in principle, be obtained in DFT by a minimization over $N$-representable one-electron densities $\denr$
\begin{equation}
 \label{eq:levy}
 E_0 = \min_{\den} \bigg\{ F[\den] + \int \d \br{} v_{\text{ne}} (\br{}) \denr \bigg\},
\end{equation}
where $v_\text{ne}(\br{})$ is the nuclei-electron potential, and $F[\den]$ is the universal Levy-Lieb density functional written with the constrained search formalism as~\cite{Lev-PNAS-79,Lie-IJQC-83}
\begin{equation}
 \label{eq:levy_func}
 F[\den] = \min_{\Psi \to \den} \mel{\Psi}{\kinop +\weeop}{\Psi},
\end{equation}
where $\kinop$ and $\weeop$ are the kinetic and electron-electron Coulomb operators, and the notation $\Psi \to \den$ means that the wave function $\Psi$ yields the density $\den$. 
The minimizing density $n_0$ in Eq.~\eqref{eq:levy} is the exact ground-state density. Nevertheless, in practical calculations, the accessible densities are necessarily restricted to the set of densities ``representable in a basis set $\Bas$'', \ie, densities coming from wave functions expandable in the $N$-electron Hilbert space generated by the one-electron basis set $\Bas$. In the following, we always consider only such representable-in-$\Bas$ densities. With this restriction, Eq.~\eqref{eq:levy} then gives an upper bound $E_0^\Bas$ of the exact ground-state energy. Since the density has a faster convergence with the size of the basis set than the wave function, this restriction is a rather weak one and we can consider that $E_0^\Bas$ is an acceptable approximation to the exact ground-state energy, \ie, $E_0^\Bas \approx E_0$.

In the present context, it is important to notice that the wave functions $\Psi$ defined in Eq.~\eqref{eq:levy_func} are not restricted to a finite basis set, \ie, they should be expanded in a complete basis set. In Ref.~\onlinecite{GinPraFerAssSavTou-JCP-18}, it was then proposed to decompose $F[\den]$ as, for a representable-in-$\Bas$ density $\den$,
\begin{equation}
 \label{eq:def_levy_bas}
 F[\den] = \min_{\wf{}{\Bas} \to \den} \mel*{\wf{}{\Bas}}{\kinop +\weeop}{\wf{}{\Bas}} + \efuncden{\den}, 
\end{equation}
where $\wf{}{\Bas}$ are wave functions expandable in the $N$-electron Hilbert space generated by $\basis$, and 
\begin{equation}
 \begin{aligned}
 \efuncden{\den} = \min_{\Psi \to \den} \mel*{\Psi}{\kinop +\weeop }{\Psi} 
                  - \min_{\Psi^{\Bas} \to \den} \mel*{\wf{}{\Bas}}{\kinop +\weeop}{\wf{}{\Bas}} 
 \end{aligned}
\end{equation}
is the complementary density functional to the basis set $\Bas$.
Introducing the decomposition in Eq.~\eqref{eq:def_levy_bas} back into Eq.~\eqref{eq:levy} yields 
\begin{multline}
 \label{eq:E0basminPsiB}
 E_0^\Bas = \min_{\Psi^{\Bas}} \bigg\{ \mel*{\wf{}{\Bas}}{\kinop +\weeop}{\wf{}{\Bas}} + \efuncden{\den_{{\Psi^{\Bas}}}} 
	\\
	+ \int \d \br{} v_{\text{ne}} (\br{}) \den_{\Psi^{\Bas}}(\br{}) \bigg\},
\end{multline}
where the minimization is only over wave functions $\wf{}{\Bas}$ restricted to the basis set $\basis$ and $\den_{{\Psi^{\Bas}}}(\br{})$ refers to the density generated from $\wf{}{\Bas}$. Therefore, thanks to Eq.~\eqref{eq:E0basminPsiB}, one can properly combine a WFT calculation in a finite basis set with a density functional (hereafter referred to as complementary functional) accounting for the correlation effects that are not included in the basis set.

As a simple non-self-consistent version of this approach, we can approximate the minimizing wave function $\Psi_0^{\Bas}$ in Eq.~\eqref{eq:E0basminPsiB} by the ground-state FCI wave function $\psifci$ within $\Bas$, and we then obtain the following approximation for the exact ground-state energy [see Eqs.~(12)--(15) of Ref.~\onlinecite{GinPraFerAssSavTou-JCP-18}]
\begin{equation}
 \label{eq:e0approx}
 E_0 \approx E_0^\Bas \approx \efci + \efuncbasisFCI,
\end{equation}
where $\efci$ and $n_\text{FCI}^\Bas$ are the ground-state FCI energy and density, respectively. As it was originally shown in Ref.~\onlinecite{GinPraFerAssSavTou-JCP-18} and further emphasized in Refs.~\onlinecite{LooPraSceTouGin-JCPL-19,GinSceTouLoo-JCP-19}, the main role of $\efuncbasisFCI$ is to correct for the basis-set incompleteness error, a large part of which originating from the lack of electron-electron cusp in the wave function expanded in an incomplete basis set. The whole purpose of this work is to determine approximations for $\efuncbasisFCI$ which are suitable for strongly correlated molecular systems. Two key requirements for this purpose are i) spin-multiplet degeneracy, and ii) size consistency.

\subsection{Effective interaction in a finite basis}
\label{sec:wee}
As originally shown by Kato, \cite{Kat-CPAM-57} the electron-electron cusp of the exact wave function originates from the divergence of the Coulomb interaction at the coalescence point. Therefore, a cuspless wave function $\wf{}{\Bas}$ could also be obtained from a Hamiltonian with a non-divergent electron-electron interaction. 
In other words, the impact of the basis set incompleteness can be understood as the removal of the divergence of the usual electron-electron Coulomb interaction.

As originally derived in Ref.~\onlinecite{GinPraFerAssSavTou-JCP-18} (see Sec.~II D~and Appendices), one can obtain an effective non-divergent electron-electron interaction, here referred to as $\wbasis$, which reproduces the expectation value of the electron-electron Coulomb interaction operator over a given wave function $\wf{}{\Bas}$. As we are interested in the behavior at the coalescence point, we focus on the opposite-spin part of the electron-electron interaction. More specifically, the effective electron-electron interaction associated to a given wave function $\wf{}{\Bas}$ is defined as
\begin{equation}
 \label{eq:wbasis}
 \wbasis =
    \begin{cases}
      \fbasis /\twodmrdiagpsi,    & \text{if $\twodmrdiagpsi \ne 0$,}
\\
       \infty,                                                                                          & \text{otherwise,}
    \end{cases}
\end{equation}
where
\begin{equation}
 \twodmrdiagpsi = \sum_{pqrs \in \Bas} \SO{p}{1} \SO{q}{2} \Gam{pq}{rs} \SO{r}{1} \SO{s}{2}, 
\end{equation}
is the opposite-spin pair density associated with $\wf{}{\Bas}$, and $\Gam{pq}{rs} = 2 \mel*{\wf{}{\Bas}}{ \aic{r_\downarrow}\aic{s_\uparrow}\ai{q_\uparrow}\ai{p_\downarrow}}{\wf{}{\Bas}}$ its associated tensor in a basis of spatial orthonormal orbitals $\{\SO{p}{}\}$, 
\begin{equation}
        \label{eq:fbasis}
        \fbasis
        = \sum_{pqrstu \in \Bas}  \SO{p}{1} \SO{q}{2} \V{pq}{rs} \Gam{rs}{tu} \SO{t}{1} \SO{u}{2},
\end{equation}
and $\V{pq}{rs}= \braket{pq}{rs}$ are the usual two-electron Coulomb integrals.
With such a definition, one can show that $\wbasis$ satisfies 
\begin{multline}
	\frac{1}{2}\iint \dr{1} \dr{2} \wbasis \twodmrdiagpsi =
	\\
	\frac{1}{2} \iint \dr{1} \dr{2} \frac{\twodmrdiagpsi}{\abs{\br{1}-\br{2}}}. 
\end{multline}
As shown in Ref.~\onlinecite{GinPraFerAssSavTou-JCP-18}, the effective interaction $\wbasis$ is necessarily finite at coalescence for an incomplete basis set, and tends to the usual Coulomb interaction in the CBS limit for any choice of wave function $\psibasis$, \ie,
\begin{equation}
 \label{eq:cbs_wbasis}
 \lim_{\Bas \to \text{CBS}} \wbasis = \frac{1}{\abs{\br{1}-\br{2}}},\quad \forall\,\psibasis. 
\end{equation}
The condition in Eq.~\eqref{eq:cbs_wbasis} is fundamental as it guarantees the correct behavior of the theory in the CBS limit.

\subsection{Local range-separation function}
\label{sec:mur}
\subsubsection{General definition}
The effective interaction within a finite basis, $\wbasis$, is bounded and resembles the long-range interaction used in RSDFT
\begin{equation}
 \label{eq:weelr}
 w_\text{ee}^{\lr}(\mu;r_{12}) = \frac{\text{erf}\big(\mu \,r_{12} \big)}{r_{12}}, 
\end{equation}
where $\mu$ is the range-separation parameter. As originally proposed in Ref.~\onlinecite{GinPraFerAssSavTou-JCP-18}, we make the correspondence between these two interactions by using the local range-separation function 
\begin{equation}
 \label{eq:def_mur}
 \murpsi = \frac{\sqrt{\pi}}{2} \wbasiscoal, 
\end{equation}
such that the two interactions coincide at the electron-electron coalescence point for each $\br{}$
\begin{equation}
 w_\text{ee}^{\lr}(\murpsi;0) = \wbasiscoal, \quad \forall \, \br{}. 
\end{equation}
Because of the very definition of $\wbasis$, one has the following property in the CBS limit [see Eq.~\eqref{eq:cbs_wbasis}]
\begin{equation}
 \label{eq:cbs_mu}
  \lim_{\Bas \to \text{CBS}} \murpsi = \infty, \quad \forall \,\psibasis, 
\end{equation}
which is again fundamental to guarantee the correct behavior of the theory in the CBS limit. 

\subsubsection{Frozen-core approximation}
\label{sec:FC}
As all WFT calculations in this work are performed within the frozen-core approximation, we use a ``valence-only'' (or no-core) version of the various quantities needed for the complementary functional introduced in Ref.~\onlinecite{LooPraSceTouGin-JCPL-19}. We partition the basis set as $\Bas = \Cor \bigcup \BasFC$, where $\Cor$ and $\BasFC$ are the sets of core and ``valence'' (\ie, non-core) orbitals, respectively, and define the valence-only local range-separation function as
\begin{equation}
 \label{eq:def_mur_val}
 \murpsival = \frac{\sqrt{\pi}}{2} \wbasiscoalval{},  
\end{equation}
where 
\begin{equation}
 \label{eq:wbasis_val}
 \wbasisval =
    \begin{cases}
      \fbasisval /\twodmrdiagpsival,    & \text{if $\twodmrdiagpsival \ne 0$,}
\\
       \infty,                                                                                          & \text{otherwise,}
    \end{cases}
\end{equation}
is the valence-only effective interaction and
\begin{gather}
	\label{eq:fbasis_val}
	\fbasisval
	= \sum_{pq\in \Bas} \sum_{rstu \in \BasFC}  \SO{p}{1} \SO{q}{2} \V{pq}{rs} \Gam{rs}{tu} \SO{t}{1} \SO{u}{2},
	\\
	\label{eq:twordm_val}
	\twodmrdiagpsival 
	= \sum_{pqrs \in \BasFC} \SO{p}{1} \SO{q}{2} \Gam{pq}{rs} \SO{r}{1} \SO{s}{2}. 
\end{gather}
One would note the restrictions of the sums to the set $\BasFC$ in Eqs.~\eqref{eq:fbasis_val} and \eqref{eq:twordm_val}. 
It is also noteworthy that, with the present definition, $\wbasisval$ still tends to the usual Coulomb interaction as $\Bas \to \CBS$. For simplicity, we will drop the indication ``val'' in the notation for the rest of the paper.

\subsection{General form of the complementary functional}
\label{sec:functional}

\subsubsection{Generic approximate form}
\label{sec:functional_form}

As originally proposed and motivated in Ref.~\onlinecite{GinPraFerAssSavTou-JCP-18}, we approximate the complementary functional $\efuncden{\den}$ by using the so-called correlation energy functional with multideterminant reference (ECMD) introduced by Toulouse \textit{et al.}\cite{TouGorSav-TCA-05,Tou-THESIS-05} Following the recent work in Ref.~\onlinecite{LooPraSceTouGin-JCPL-19}, we propose to consider a Perdew-Burke-Ernzerhof (PBE)-like functional which uses the one-electron density $\denr$, the spin polarization $\zeta(\br{})=[n_\uparrow(\br{})-n_\downarrow(\br{})]/\denr$ (where $n_\uparrow(\br{})$ and $n_\downarrow(\br{})$ are the spin-up and spin-down densities), the reduced density gradient $s(\br{}) = \nabla \denr/\denr^{4/3}$, and the on-top pair density $\ntwo(\br{})\equiv \ntwo(\br{},\br{})$. In the present work, all these quantities are computed with the same wave function $\psibasis$ used to define $\mur \equiv\murpsi$. 
Therefore, $\efuncden{\den}$ has the following generic form
\begin{multline}
 \label{eq:def_ecmdpbebasis}
 \efuncdenpbe{\argebasis} =  
 \\ 
 \int  \d\br{} \,\denr   \ecmd(\argrebasis),
\end{multline}
where 
\begin{equation}
 \label{eq:def_ecmdpbe}
 \ecmd(\argecmd) = \frac{\varepsilon_{\text{c}}^{\text{PBE}}(\argepbe)}{1+ \beta(\argepbe,\ntwo) \; \mu^3}
\end{equation}
is the correlation energy per particle, with 
\begin{equation}
 \label{eq:def_beta}
 \beta(\argepbe,\ntwo) = \frac{3}{2\sqrt{\pi}(1 - \sqrt{2})}\frac{\varepsilon_{\text{c}}^{\text{PBE}}(\argepbe)}{\ntwo/\den}, 
\end{equation}
where $\varepsilon_{\text{c}}^{\text{PBE}}(\argepbe)$ is the usual PBE correlation energy per particle. \cite{PerBurErn-PRL-96} Before introducing the different flavors of approximate functionals that we will use here (see Sec.~\ref{sec:def_func}), we would like to give some motivations for this choice of functional form. 

The form of $\ecmd(\argecmd)$ in Eq.~\eqref{eq:def_ecmdpbe} has been originally proposed in Ref.~\onlinecite{FerGinTou-JCP-18} in the context of RSDFT. In the $\mu\to 0$ limit, it reduces to the usual PBE correlation functional, \ie,
\begin{equation}
   \lim_{\mu \to 0} \ecmd(\argecmd) = \varepsilon_{\text{c}}^{\text{PBE}}(\argepbe),   
\end{equation}
which is relevant in the weak-correlation (or high-density) limit. In the large-$\mu$ limit, it behaves as
\begin{equation}
 \label{eq:lim_mularge}
    \ecmd(\argecmd) \isEquivTo{\mu\to\infty} \frac{2\sqrt{\pi}(1 - \sqrt{2})}{3 \mu^3}  \frac{\ntwo}{n},
\end{equation}
which is the exact large-$\mu$ behavior of the exact ECMD correlation energy. \cite{PazMorGorBac-PRB-06,FerGinTou-JCP-18} Of course, for a specific system, the large-$\mu$ behavior will be exact only if one uses for $n_2$ the \textit{exact} on-top pair density of this system. This large-$\mu$ limit in Eq.~\eqref{eq:lim_mularge} is relevant in the strong-correlation (or low-density) limit. In the context of RSDFT, some of the present authors have illustrated in Ref.~\onlinecite{FerGinTou-JCP-18} that the on-top pair density involved in Eq.~\eqref{eq:def_ecmdpbe} plays indeed a crucial role when reaching the strong-correlation regime. The importance of the on-top pair density in the strong-correlation regime have been also recently acknowledged by Gagliardi and coworkers \cite{CarTruGag-JPCA-17} and Pernal and coworkers.\cite{GritMeePer-PRA-18}

Note also that $\ecmd(\argecmd)$ vanishes when $\ntwo$ vanishes, \ie,
\begin{equation}
 \label{eq:lim_n2}
  \lim_{\ntwo \to 0} \ecmd(\argecmd) = 0,
\end{equation}
which is expected for systems with a vanishing on-top pair density.
Finally, the function $\ecmd(\argecmd)$ vanishes when $\mu \to \infty$ like all RSDFT short-range functionals, \ie,
\begin{equation}
 \label{eq:lim_muinf}
  \lim_{\mu \to \infty} \ecmd(\argecmd) = 0. 
\end{equation}

\subsubsection{Two limits where the complementary functional vanishes}
\label{sec:functional_prop}

Within the definitions of Eqs.~\eqref{eq:def_mur} and \eqref{eq:def_ecmdpbebasis}, any approximate complementary functional $\efuncdenpbe{\argebasis}$ satisfies two important properties.

First, thanks to the properties in Eqs.~\eqref{eq:cbs_mu} and~\eqref{eq:lim_muinf}, $\efuncdenpbe{\argebasis}$ vanishes in the CBS limit, independently of the type of wave function $\psibasis$ used to define the local range-separation function $\mu(\br{})$ in a given basis set $\Bas$, 
\begin{equation}
 \label{eq:lim_ebasis}
 \lim_{\basis \to \text{CBS}} \efuncdenpbe{\argebasis} = 0, \quad \forall\, \psibasis. 
\end{equation}

Second, $\efuncdenpbe{\argebasis}$ correctly vanishes for systems with uniformly vanishing on-top pair density, such as one-electron systems and for the stretched H$_2$ molecule, 
\begin{equation}
 \lim_{n_2 \to 0} \efuncdenpbe{\argebasis} = 0.
\end{equation}
This property is doubly guaranteed by i) the choice of setting $\wbasis = \infty$ for a vanishing pair density [see Eq.~\eqref{eq:wbasis}], which leads to $\mu(\br{}) \to \infty$ and thus a vanishing $\ecmd(\argecmd)$ [see Eq.~\eqref{eq:lim_muinf}], and ii) the fact that $\ecmd(\argecmd)$ vanishes anyway when the on-top pair density vanishes [see Eq.~\eqref{eq:lim_n2}]. 

\subsection{Requirements on the complementary functional for strong correlation}
\label{sec:requirements}

An important requirement for any electronic-structure method is size consistency, \ie, the additivity of the energies of non-interacting fragments, which is mandatory to avoid any ambiguity in computing interaction energies. When two subsystems \ce{A} and \ce{B} dissociate in closed-shell systems, as in the case of weak intermolecular interactions for instance, spin-restricted Hartree-Fock (RHF) is size-consistent. When the two subsystems dissociate in open-shell systems, such as in covalent bond breaking, it is well known that the RHF approach fails and an alternative is to use a complete-active-space self-consistent-field (CASSCF) wave function which, provided that the active space has been properly chosen, leads to additive energies. 

Another important requirement is spin-multiplet degeneracy, \ie, the independence of the energy with respect to the $S_z$ component of a given spin state, which is also a property of any exact wave function. Such a property is also important in the context of covalent bond breaking where the ground state of the supersystem $\ce{A + B}$ is generally of lower spin than the corresponding ground states of the fragments (\ce{A} and \ce{B}) which can have multiple $S_z$ components.

\subsubsection{Spin-multiplet degeneracy}

A sufficient condition to achieve spin-multiplet degeneracy is to eliminate all dependencies on $S_z$. In the case of the function $\ecmd(\argecmd)$, this means removing the dependence on the spin polarization $\zeta(\br{})$ originating from the PBE correlation functional $\varepsilon_{\text{c}}^{\text{PBE}}(\argepbe)$ [see Eq.~\eqref{eq:def_ecmdpbe}]. 

To do so, it has been proposed to replace the dependence on the spin polarization by the dependence on the on-top pair density. Most often, it is done by introducing an effective spin polarization~\cite{MosSan-PRA-91,BecSavSto-TCA-95,Sav-INC-96a,Sav-INC-96,MieStoSav-MP-97,TakYamYam-CPL-02,TakYamYam-IJQC-04,GraCre-MP-05,TsuScuSav-JCP-10,LimCarLuoMaOlsTruGag-JCTC-14,GarBulHenScu-JCP-15,GarBulHenScu-PCCP-15,CarTruGag-JCTC-15,GagTruLiCarHoyBa-ACR-17} (see, also, Refs.~\onlinecite{PerSavBur-PRA-95,StaDav-CPL-01})
\begin{equation}
 \label{eq:def_effspin}
 \tilde{\zeta}(n,n_{2}) =  \sqrt{ 1 - 2 \; n_{2}/n^2 }
\end{equation}
expressed as a function of the density $n$ and the on-top pair density $n_2$ calculated from a given wave function. The advantage of this approach is that this effective spin  polarization $\tilde{\zeta}$ is independent from $S_z$ since the on-top pair density is $S_z$-independent. Nevertheless, the use of $\tilde{\zeta}$ in Eq.~\eqref{eq:def_effspin} presents some disadvantages since this expression was derived for a single-determinant wave function. Hence, it does not appear justified to use it for a multideterminant wave function. More particularly, it may happen, in the multideterminant case, that $1 - 2 \; n_{2}/n^2 < 0 $ which results in a complex-valued effective spin polarization. \cite{BecSavSto-TCA-95}
Therefore, following other authors, \cite{MieStoSav-MP-97,LimCarLuoMaOlsTruGag-JCTC-14,GarBulHenScu-JCP-15} we use the following definition 
\begin{equation}
 \label{eq:def_effspin-0}
	\tilde{\zeta}(n,n_{2}) =  
   \begin{cases}
		\sqrt{ 1 - 2 \; n_{2}/n^2 },	&	\text{if } n^2 \ge 2 n_{2},
		\\
		0,								& \text{otherwise.}
   \end{cases}
\end{equation}

An alternative way to eliminate the $S_z$ dependence is to simply set $\zeta=0$, \ie, to resort to the spin-unpolarized functional. This lowers the accuracy for open-shell systems at $\mu=0$, \ie, for the usual PBE correlation functional $\varepsilon_{\text{c}}^{\text{PBE}}(\argepbe)$. Nevertheless, we argue that, for sufficiently large $\mu$, it is a viable option. Indeed, the purpose of introducing the spin polarization in semilocal density-functional approximations is to mimic the exact on-top pair density, \cite{PerSavBur-PRA-95} but our functional $\ecmd(\argecmd)$ already explicitly depends on the on-top pair density [see Eqs.~\eqref{eq:def_ecmdpbe} and \eqref{eq:def_beta}]. The dependencies on $\zeta$ and $n_2$ can thus be expected to be largely redundant. Consequently, we propose here to test the use of $\ecmd$ with \textit{a zero spin polarization}. This ensures its $S_z$ independence and, as will be numerically demonstrated, very weakly affects the complementary functional accuracy.

\subsubsection{Size consistency}

Since $\efuncdenpbe{\argebasis}$ is computed via a single integral over $\mathbb{R}^3$ [see Eq.~\eqref{eq:def_ecmdpbebasis}] which involves only local quantities [$n(\br{})$, $\zeta(\br{})$, $s(\br{})$, $n_2(\br{})$, and $\mu(\br{})$], in the case of non-overlapping fragments $\text{A}+\text{B}$, it can be written as the sum of two local contributions: one coming from the integration over the region of subsystem \ce{A} and the other one from the region of subsystem \ce{B}. Therefore, a sufficient condition for size consistency is that these quantities locally coincide in the isolated fragments and in the supersystem $\text{A}+\text{B}$. Since these local quantities are calculated from the wave function $\psibasis$, a sufficient condition is that the wave function is multiplicatively separable in the limit of non-interacting fragments, \ie, $\ket*{\Psi_{\text{A}+\text{B}}^{\basis}} = \ket*{\Psi_{\ce{A}}^{\basis}} \otimes \ket*{\Psi_{\ce{B}}^{\basis}}$. We refer the interested reader to Appendix~\ref{app:sizeconsistency} for a detailed proof and discussion of the latter statement. 
In the case where the two subsystems \ce{A} and \ce{B} dissociate in closed-shell systems, a simple RHF wave function ensures this property, but when one or several covalent bonds are broken, a properly chosen CASSCF wave function can be used to recover this property. The underlying active space must however be chosen in such a way that it leads to size-consistent energies in the limit of dissociated fragments.

\subsection{Actual approximations used for the complementary functional}
\label{sec:def_func}

As the present work focuses on the strong-correlation regime, we propose here to investigate only approximate functionals which are $S_z$ independent and size-consistent in the case of covalent bond breaking. Therefore, the wave functions $\psibasis$ used throughout this paper are CASSCF wave functions in order to ensure size consistency of all local quantities. The difference between the different flavors of functionals are only due to the types of spin polarization and on-top pair density used.

Regarding the spin polarization that enters into the function $\varepsilon_{\text{c}}^{\text{PBE}}(\argepbe)$, two different types of $S_z$-independent formulations are considered: i) the \textit{effective} spin polarization $\tilde{\zeta}$ defined in Eq.~\eqref{eq:def_effspin-0} and calculated from the CASSCF wave function, and ii) a \textit{zero} spin polarization. In the latter case, the functional is referred as to ``SU'' which stands for ``spin unpolarized''. 

Regarding the on-top pair density entering in Eq.~\eqref{eq:def_beta}, we use two different approximations. The first one is based on the uniform electron gas (UEG) and reads 
\begin{equation}
 \label{eq:def_n2ueg}
 \ntwo^{\text{UEG}}(n,\zeta) \approx n^2\big(1-\zeta^2\big)g_0(n),
\end{equation}
where the pair-distribution function $g_0(n)$ is taken from Eq.~(46) of Ref.~\onlinecite{GorSav-PRA-06}. As the spin polarization appears in Eq.~\eqref{eq:def_n2ueg}, we use the effective spin polarization $\tilde{\zeta}$ of Eq.~\eqref{eq:def_effspin-0} in order to ensure $S_z$ independence. Thus, $\ntwo^{\text{UEG}}$ will depend indirectly on the on-top pair density of the CASSCF wave function through $\tilde{\zeta}$. When using $\ntwo^{\text{UEG}}(\br{}) \equiv \ntwo^{\text{UEG}}(n(\br{}),\tilde{\zeta}(\br{}))$ in a functional, we will refer to it as ``UEG''.

The second approach to approximate the exact on-top pair density consists in using directly the on-top pair density of the CASSCF wave function. Following the work of some of the present authors, \cite{FerGinTou-JCP-18,GinSceTouLoo-JCP-19} we introduce the extrapolated on-top pair density
\begin{equation}
 \label{eq:def_n2extrap}
 \ntwoextrap(\ntwo,\mu) = \qty( 1 + \frac{2}{\sqrt{\pi}\mu} )^{-1} \; \ntwo,
\end{equation}
which directly follows from the large-$\mu$ extrapolation of the exact on-top pair density derived by Gori-Giorgi and Savin\cite{GorSav-PRA-06} in the context of RSDFT. Thus, the extrapolated on-top pair density $\ntwoextrap$ depends on the local range-separation function $\mu$. When using $\ntwoextrap(\br{}) \equiv \ntwoextrap(\ntwo(\br{}),\mu(\br{}))$ in a functional, we will simply refer it as ``OT'', which stands for "on-top". 

We then define three complementary functionals:
\begin{itemize}

\item[i)] $\pbeuegXi$ which combines the effective spin polarization of Eq.~\eqref{eq:def_effspin-0}  and  the UEG on-top pair density defined in Eq.~\eqref{eq:def_n2ueg}:
\begin{multline}
	\label{eq:def_pbeueg_i}
	\bar{E}^\Bas_{\pbeuegXi} 
	\\
	=  \int  \d\br{} \,\denr  \ecmd(\argrpbeuegXi), 
\end{multline}
\item[ii)] $\pbeontXi$ which combines the effective spin polarization of Eq.~\eqref{eq:def_effspin-0} and the on-top pair density of Eq.~\eqref{eq:def_n2extrap}:
\begin{equation}
	\label{eq:def_pbeueg_ii}
	\bar{E}^\Bas_{\pbeontXi} =  \int  \d\br{} \,\denr  \ecmd(\argrpbeontXi),  
\end{equation}
\item[iii)] $\pbeontns$ which combines a zero spin polarization and the on-top pair density of Eq.~\eqref{eq:def_n2extrap}:
\begin{equation}
	\label{eq:def_pbeueg_iv}
	\bar{E}^\Bas_{\pbeontns}  =  \int  \d\br{} \,\denr  \ecmd(\argrpbeontns).  
\end{equation}
\end{itemize}
The performance of each of these functionals is tested in the following. Note that we did not define a spin-unpolarized version of the PBE-UEG functional because it would have been significantly inferior (in terms of performance) compared to the three other functionals. Indeed, because to the lack of knowledge on the spin polarization or on the accurate on-top pair density, such a functional would be inaccurate in particular for open-shell systems. This assumption has been numerically confirmed by calculations.

%%%%%%%%%%%%%%%%%%%%%%%%
\section{Results}
\label{sec:results}

\subsection{Computational details}

We present potential energy curves of small molecules up to the dissociation limit
to investigate the performance of the basis-set correction in regimes of both weak and strong correlation.
The considered systems are the \ce{H10} linear chain with equally-spaced atoms, and the \ce{N2}, \ce{O2}, and \ce{F2} diatomics.

The computation of the ground-state energy in Eq.~\eqref{eq:e0approx} in a given basis set requires approximations to the FCI energy $\efci$ and to the basis-set correction $\efuncbasisFCI$. 
For diatomics with the aug-cc-pVDZ and aug-cc-pVTZ basis sets,~\cite{KenDunHar-JCP-92} energies are obtained using frozen-core selected-CI calculations (using the CIPSI algorithm) followed by the extrapolation scheme proposed by Holmes \textit{et al.} (see Refs.~\onlinecite{HolUmrSha-JCP-17, SceGarCafLoo-JCTC-18, LooSceBloGarCafJac-JCTC-18, SceBenJacCafLoo-JCP-18, LooBogSceCafJac-JCTC-19, QP2} for more detail). All these calculations are performed with the latest version of \textsc{Quantum Package}, \cite{QP2} and will be labeled as exFCI in the following. In the case of \ce{F2}, we also use the correlation energy extrapolated by intrinsic scaling (CEEIS) \cite{BytNagGorRue-JCP-07} as an estimate of the FCI correlation energy with the cc-pVXZ (X $=$ D, T, and Q) basis sets.~\cite{Dun-JCP-89} The estimated exact potential energy curves are obtained from experimental data \cite{LieCle-JCP-74a} for the \ce{N2} and \ce{O2} molecules, and from CEEIS calculations in the case of \ce{F2}. For all geometries and basis sets, the error with respect to the exact FCI energies are estimated to be of the order of $0.5$~mHa.
For the three diatomics, we performed an additional exFCI calculation with the aug-cc-pVQZ basis set at the equilibrium geometry to obtain reliable estimates of the FCI/CBS dissociation energy.
In the case of the \ce{H10} chain, the approximation to the FCI energies together with the estimated exact potential energy curves are obtained from the data of Ref.~\onlinecite{h10_prx} where the authors performed MRCI+Q calculations with a minimal valence active space as reference (see below for the description of the active space). 

We note that, even though we use near-FCI energies in this work, the DFT-based basis-set correction could also be applied to any approximation to FCI such as multireference perturbation theory, similarly to what was done for weakly correlated systems for which the basis-set correction was applied to CCSD(T) calculations. \cite{LooPraSceTouGin-JCPL-19}

Regarding the complementary functional, we first perform full-valence CASSCF calculations with the GAMESS-US software~\cite{gamess} to obtain the wave function $\psibasis$. Then, all density-related quantities involved in the functional [density $n(\br{})$, effective spin polarization $\tilde{\zeta}(\br{})$, reduced density gradient $s(\br{})$, and on-top pair density $n_2(\br{})$] together with the local range-separation function $\mu(\br{})$ are calculated with this full-valence CASSCF wave function. The CASSCF calculations are performed with the following active spaces: (10e,10o) for \ce{H10}, (10e,8o) for \ce{N2}, (12e,8o) for \ce{O2}, and (14e,8o) for \ce{F2}. We note that, instead of using CASSCF wave functions for $\psibasis$, one could of course use the same selected-CI wave functions used for calculating the energy but the calculations of $n_2(\br{})$ and $\mu(\br{})$ would then be more costly. 

Also, as the frozen-core approximation is used in all our selected-CI calculations, we use the corresponding valence-only complementary functionals (see Subsec.~\ref{sec:FC}). Therefore, all density-related quantities exclude any contribution from the 1s core orbitals, and the range-separation function follows the definition given in Eq.~\eqref{eq:def_mur_val}. 

It should be stressed that the computational cost of the basis-set correction (see Appendix~\ref{app:computational}) is negligible compared to the cost the selected-CI calculations.

\subsection{H$_{10}$ chain}

The \ce{H10} chain with equally-spaced atoms is a prototype of strongly correlated systems as it consists in the simultaneous breaking of 10 interacting covalent $\sigma$ bonds.
As it is a relatively small system, benchmark calculations at near-CBS values are available (see Ref.~\onlinecite{h10_prx} for a detailed study of this system). 

%%%%%%%%%%%%%%%%%%%%%%%%%%%%%%%%%%%%%%%%%%%%%%%%%%%%%%%%%%%%%%%%%%%%%%%%%%%%%%%%%%%%%%%%%%%%%%%%%%%%%%%%%%%%%%%%
\begin{figure*}
	\subfigure[cc-pVDZ]{
        \includegraphics[width=0.45\linewidth]{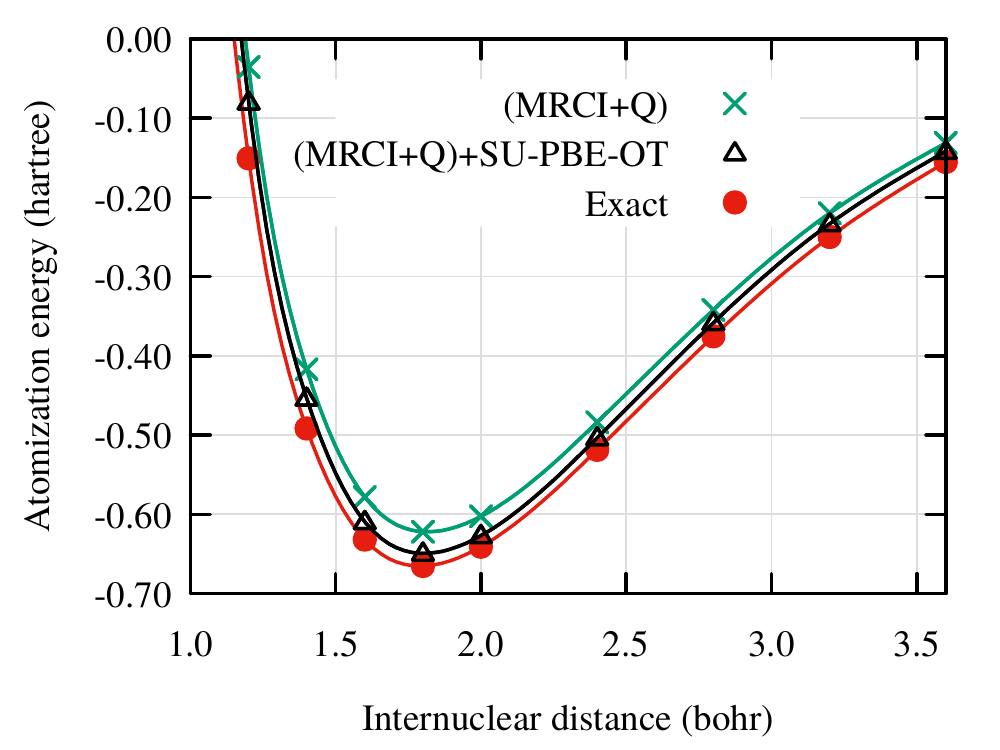}
        \includegraphics[width=0.45\linewidth]{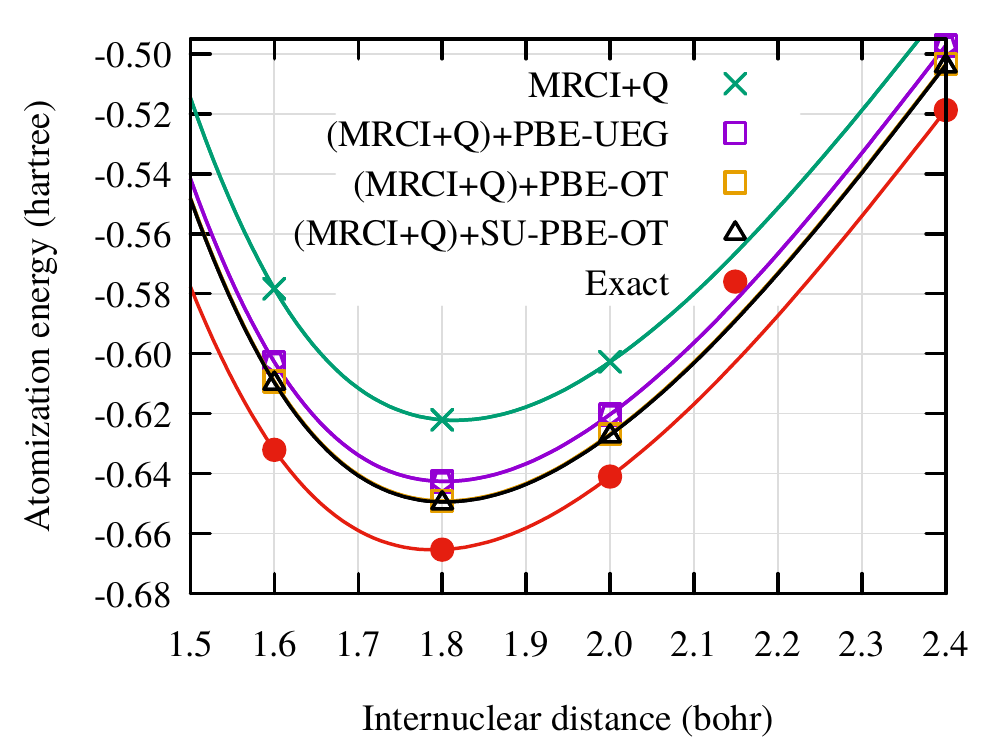}
       }
	\subfigure[cc-pVTZ]{
        \includegraphics[width=0.45\linewidth]{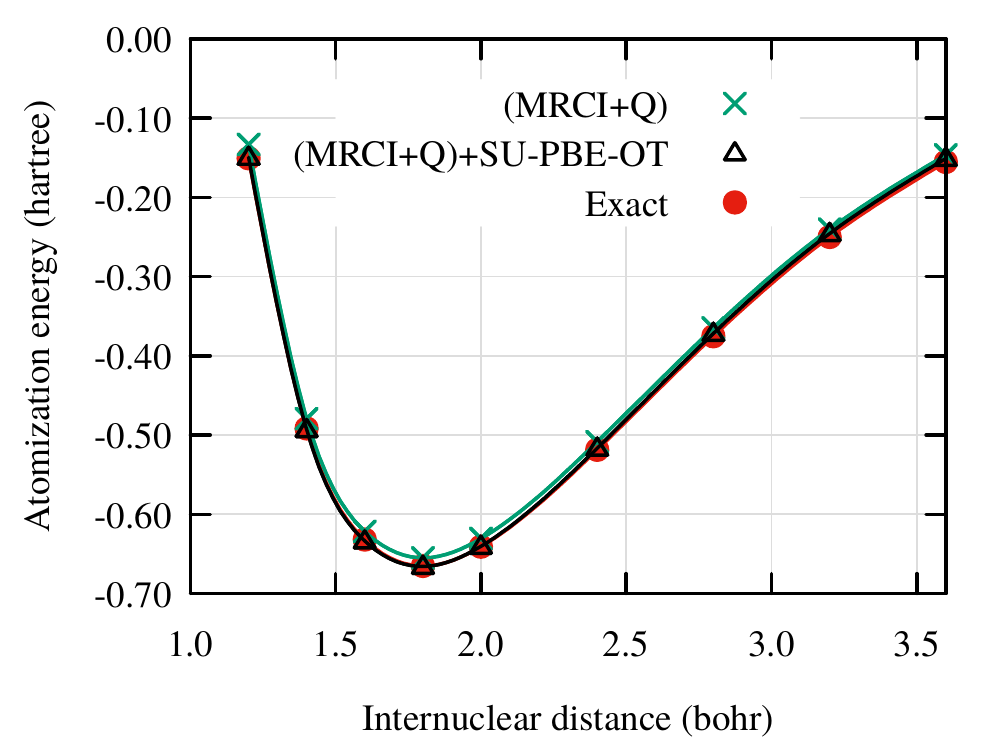}
        \includegraphics[width=0.45\linewidth]{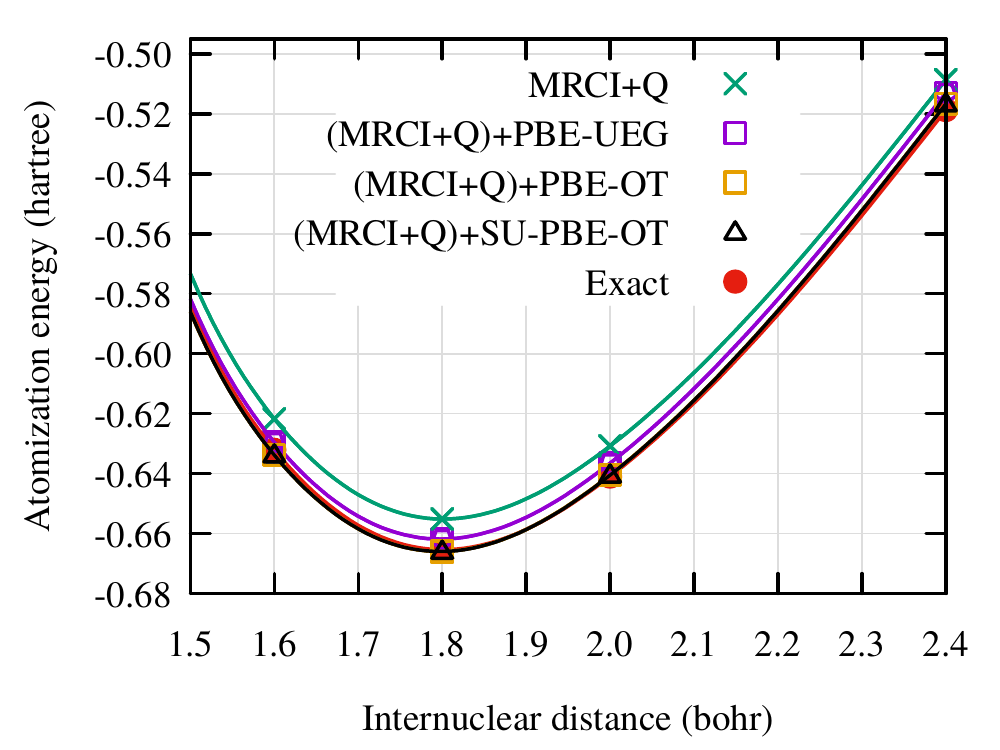}
       }
	\subfigure[cc-pVQZ]{
        \includegraphics[width=0.45\linewidth]{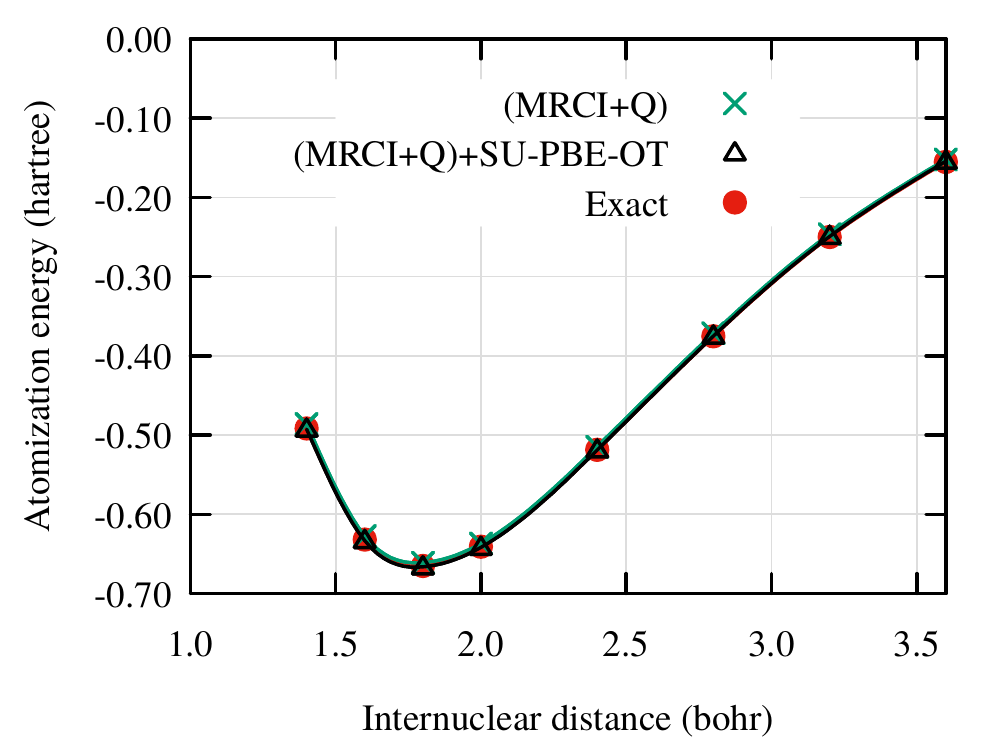}
        \includegraphics[width=0.45\linewidth]{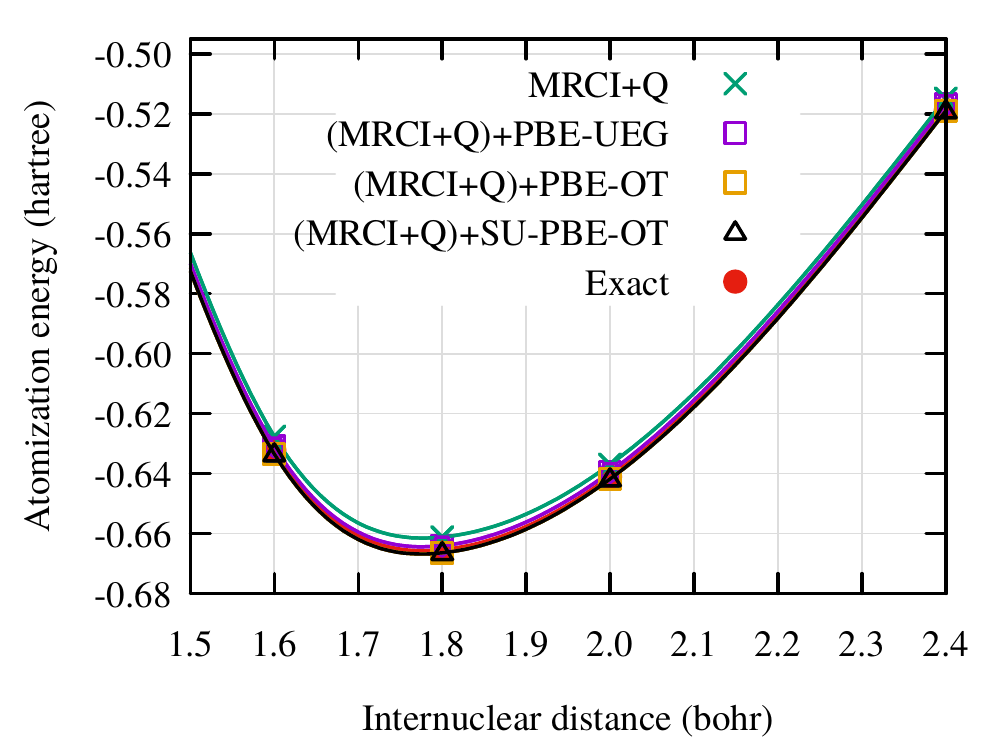}
       }
        \caption{
        Potential energy curves of the H$_{10}$ chain with equally-spaced atoms calculated with MRCI+Q and basis-set corrected MRCI+Q using the cc-pVDZ (top), cc-pVTZ (center), and cc-pVQZ (bottom) basis sets.
        The MRCI+Q energies and the estimated exact energies have been extracted from Ref.~\onlinecite{h10_prx}.
        \label{fig:H10}}
\end{figure*}
%%%%%%%%%%%%%%%%%%%%%%%%%%%%%%%%%%%%%%%%%%%%%%%%%%%%%%%%%%%%%%%%%%%%%%%%%%%%%%%%%%%%%%%%%%%%%%%%%%%%%%%%%%%%%%%%

%%%%%%%%%%%%%%%%%%%%%%%%%%%%%%%%%%%%%%%%%%%%%%%%%%%%%%%%%%%%%%%%%%%%%%%%%%%%%%%%%%%%%%%%%%%%%%%%%%%%%%%%%%%%%%%%%%%%%%%%%%%%%%%%%
\begin{table*}
\caption{Atomization energies (in mHa) and associated errors (in square brackets) with respect to the estimated exact values computed at different levels of theory with various basis sets.}
\begin{ruledtabular}
\begin{tabular}{lrdddd}

   System 	&  \tabc{Basis set}        &  \tabc{MRCI+Q\fnm[1]}              &   \tabc{(MRCI+Q)+$\pbeuegXi$}   &   \tabc{(MRCI+Q)+$\pbeontXi$}   &     \tabc{(MRCI+Q)+$\pbeontns$}    \\
\hline                                                                                                                 
\ce{H10} & cc-pVDZ           &  622.1 [43.3]            &   642.6 [22.8]         &   649.2  [16.2]          &     649.5  [15.9]           \\
         & cc-pVTZ           &  655.2 [10.2]            &   661.9 [3.5]          &   666.0  [-0.6]          &     666.0  [-0.6]           \\
         & cc-pVQZ           &  661.2 [4.2]             &   664.1 [1.3]          &   666.4  [-1.0]          &     666.5  [-1.1]           \\[0.1cm]
  	&	&\multicolumn{4}{c}{Estimated exact:\fnm[1] 665.4} \\[0.2cm]
\hline
    		&         		& \tabc{exFCI}                    &   \tabc{exFCI+$\pbeuegXi$}   &   \tabc{exFCI+$\pbeontXi$}   &     \tabc{exFCI+$\pbeontns$}\\
\hline                                                                                                                 
\ce{N2} & aug-cc-pVDZ          &   321.9  [40.8]         &   356.2  [6.5]           &   355.5  [7.2]           &     354.6  [8.1]           \\
        & aug-cc-pVTZ          &   348.5  [14.2]         &   361.5  [1.2]           &   363.5  [-0.5]          &     363.2  [-0.3]           \\
        & aug-cc-pVQZ          &   356.6  [6.1 ]         &   362.8  [-0.1]          &   364.2  [-1.5]          &     364.3  [-1.6]           \\[0.1cm]
 	&	 & \multicolumn{4}{c}{Estimated exact:\fnm[2] 362.7} \\[0.2cm]
\hline                                                                                                                 
    		&         		& \tabc{exFCI}                    &   \tabc{exFCI+$\pbeuegXi$}   &   \tabc{exFCI+$\pbeontXi$}   &     \tabc{exFCI+$\pbeontns$}\\
\hline                                                                                                                 
\ce{O2} & aug-cc-pVDZ          &   171.4  [20.5]         &   187.6  [4.3]      &   187.6  [4.3]      &     187.1  [4.8]           \\
        & aug-cc-pVTZ          &   184.5  [7.4]          &   190.3  [1.6]      &   191.2  [0.7]      &     191.0  [0.9]            \\
        & aug-cc-pVQZ          &   188.3  [3.6]          &   190.3  [1.6]      &   191.0  [0.9]      &     190.9  [1.0]            \\[0.1cm]
 	&	 & \multicolumn{4}{c}{Estimated exact:\fnm[2] 191.9} \\[0.2cm]
\hline                                                                                                                 
    		&         		& \tabc{exFCI}                    &   \tabc{exFCI+$\pbeuegXi$}   &   \tabc{exFCI+$\pbeontXi$}   &     \tabc{exFCI+$\pbeontns$}\\
\hline                                                                                                                 
\ce{F2} & aug-cc-pVDZ          &   49.6  [12.6]          &    54.8  [7.4]      &   54.9  [7.3]       &     54.8  [7.4]             \\
        & aug-cc-pVTZ          &   59.3  [2.9]           &    61.2  [1.0]      &   61.5  [0.7]       &     61.5  [0.7]             \\
        & aug-cc-pVQZ          &   60.1  [2.1]           &    61.0  [1.2]      &   61.3  [0.9]       &     61.3  [0.9]             \\[0.1cm]
        \cline{2-6}\\[-0.3cm]
    		&         		& \tabc{CEEIS\fnm[3]}                    &   \tabc{CEEIS\fnm[3]+$\pbeuegXi$}   &   \tabc{CEEIS\fnm[3]+$\pbeontXi$}   &     \tabc{CEEIS\fnm[3]+$\pbeontns$}\\
        \cline{2-6}\\[-0.3cm]
        &     cc-pVDZ          &   43.7  [18.5]          &    51.0  [11.2]     &   51.0 [11.2]       &     50.7  [11.5]            \\
        &     cc-pVTZ          &   56.3  [5.9]           &    59.2  [3.0]      &   59.6 [2.6]        &     59.5  [2.7]             \\
        &     cc-pVQZ          &   59.9  [2.3]           &    61.3  [0.9]      &   61.6 [0.6]        &     61.6  [0.6]             \\[0.1cm]
 	&	& \multicolumn{4}{c}{Estimated exact:\fnm[2] 62.2} \\
\end{tabular}
\end{ruledtabular}
\fnt[1]{From Ref.~\onlinecite{h10_prx}.}
\fnt[2]{From the CEEIS valence-only non-relativistic calculations of Ref.~\onlinecite{BytLaiRuedenJCP05}.}
\fnt[3]{From the CEEIS valence-only non-relativistic calculations of Ref.~\onlinecite{BytNagGorRue-JCP-07}.}
\label{tab:d0}
\end{table*}
%%%%%%%%%%%%%%%%%%%%%%%%%%%%%%%%%%%%%%%%%%%%%%%%%%%%%%%%%%%%%%%%%%%%%%%%%%%%%%%%%%%%%%%%%%%%%%%%%%%%%%%%%%%%%%%%%%%%%%%%%%%%%%%%%

We report in Fig.~\ref{fig:H10} the potential energy curves computed using the cc-pVXZ (X $=$ D, T, and Q) basis sets for different levels of approximation, and the corresponding  atomization energies are reported in Table \ref{tab:d0}.
As a general trend, the addition of the basis-set correction globally improves
the quality of the potential energy curves, independently of the approximation level of $\efuncbasis$. Also, no erratic behavior is found when stretching the bonds, which shows that the present procedure (\ie, the determination of the range-separation function and the definition of the functionals) is robust when reaching the strong-correlation regime.  
In other words, smooth potential energy curves are obtained with the present basis-set correction.
More quantitatively, the values of the atomization energies are within chemical accuracy (\ie, an error below $1.4$ mHa) with the cc-pVTZ basis set when using the $\pbeontXi$ and $\pbeontns$ functionals, whereas such an accuracy is not yet reached at the standard MRCI+Q/cc-pVQZ level of theory. 

Analyzing more carefully the performance of the different types of approximate functionals, the results show that $\pbeontXi$ and $\pbeontns$ are very similar (the maximal difference on the atomization energy being 0.3 mHa), and that they give slightly more accurate results than $\pbeuegXi$. These findings provide two important clues on the role of the different physical ingredients included in these functionals: i) the explicit use of the on-top pair density originating from the CASSCF wave function [see Eq.~\eqref{eq:def_n2extrap}] is preferable over the use of the UEG on-top pair density [see Eq.~\eqref{eq:def_n2ueg}] which is somewhat understandable, and ii) removing the dependence on any kind of spin polarization does not lead to a significant loss of accuracy providing that one employs a qualitatively correct on-top pair density. The latter point is crucial as it confirms that the spin polarization in density-functional approximations essentially plays the same role as the on-top pair density. This could have significant implications for the construction of more robust families of density-functional approximations within DFT.

\subsection{Dissociation of diatomics}

The \ce{N2}, \ce{O2} and \ce{F2} molecules are complementary to the \ce{H10} system for the present study. The level of strong correlation in these diatomics also increases while stretching the bonds, similarly to the case of \ce{H10}, but in addition these molecules exhibit more important and versatile types of weak correlations due to the larger number of electrons. Indeed, the short-range correlation effects are known to play a strong differential effect on the computation of the atomization energy at equilibrium, while the shape of the curve far from the equilibrium geometry is governed by dispersion interactions which are medium to long-range weak-correlation effects. \cite{AngDobJanGou-BOOK-20} The dispersion interactions in \ce{H10} play a minor role on the potential energy curve due to the much smaller number of near-neighbor electron pairs compared to  \ce{N2}, \ce{O2} or \ce{F2}. Also, \ce{O2} has a triplet ground state and is therefore a good candidate for checking the spin-polarization dependence of the various functionals proposed here.

%%%%%%%%%%%%%%%%%%%%%%%%%%%%%%%%%%%%%%%%%%%%%%%%%%%%%%%%%%%%%%%%%%%%%%%%%%%%%%%%%%%%%%%%%%%%%%%%%%%%%%%%%%%%%%%%%%%%%%%%%%%%%%%%%
\begin{figure*}
	\subfigure[aug-cc-pVDZ]{
        \includegraphics[width=0.45\linewidth]{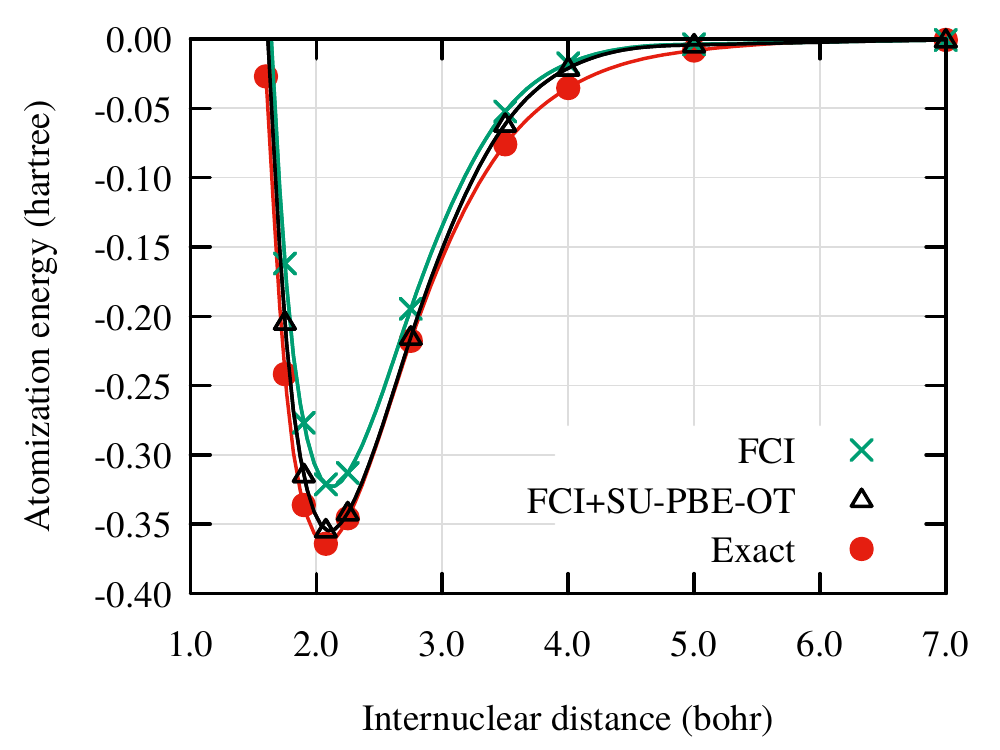}
        \includegraphics[width=0.45\linewidth]{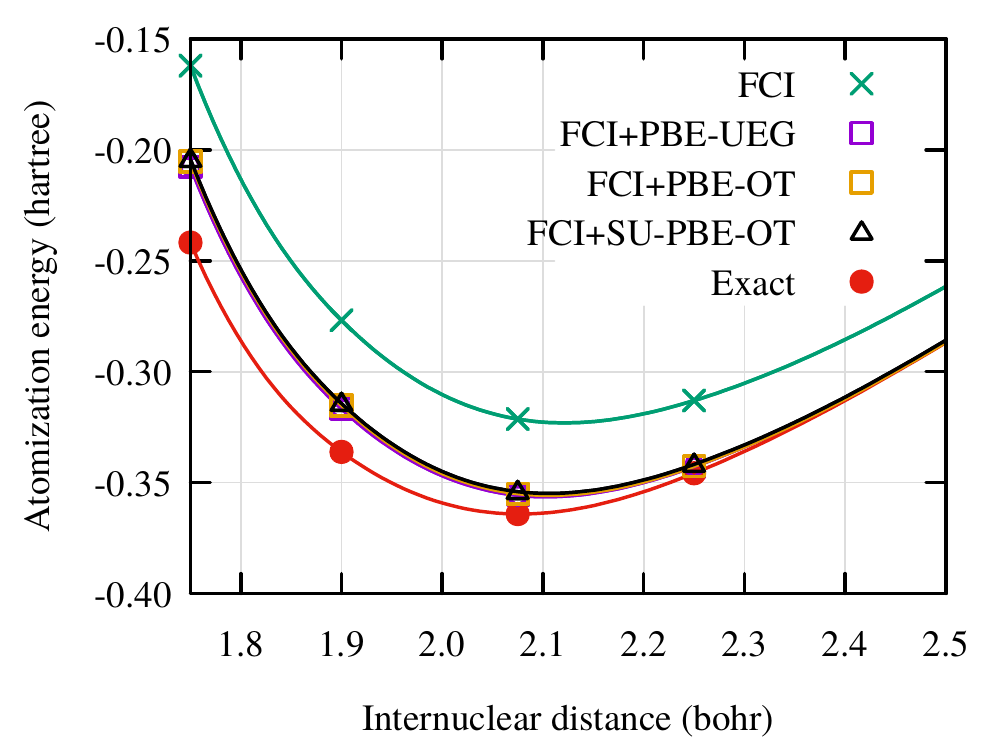}
     }
	\subfigure[aug-cc-pVTZ]{
        \includegraphics[width=0.45\linewidth]{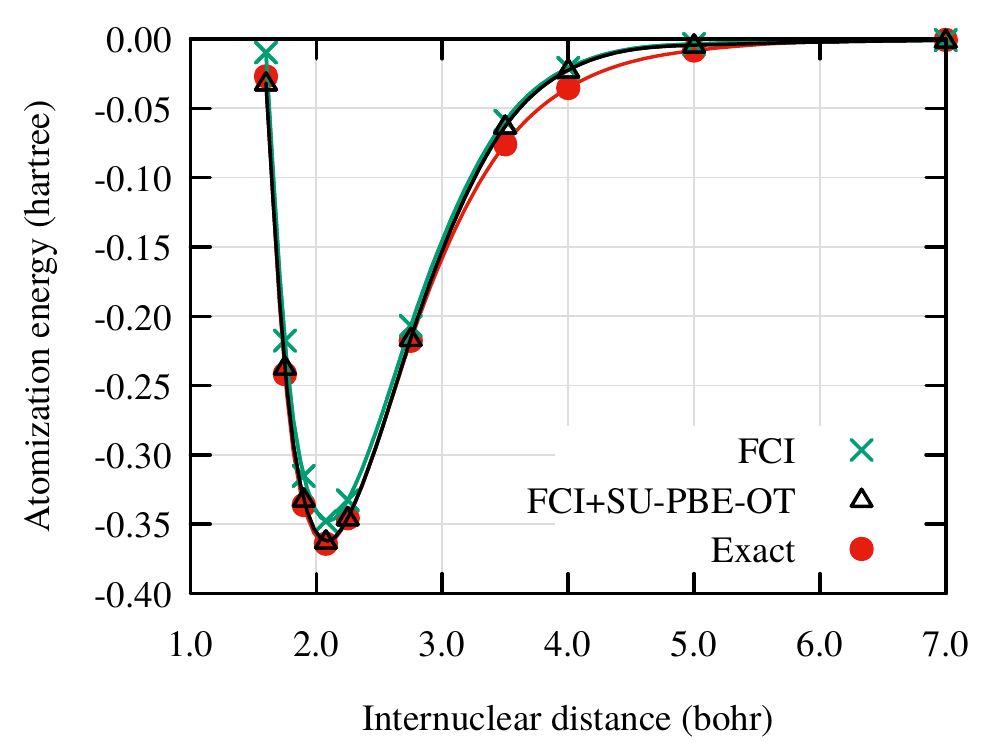}
        \includegraphics[width=0.45\linewidth]{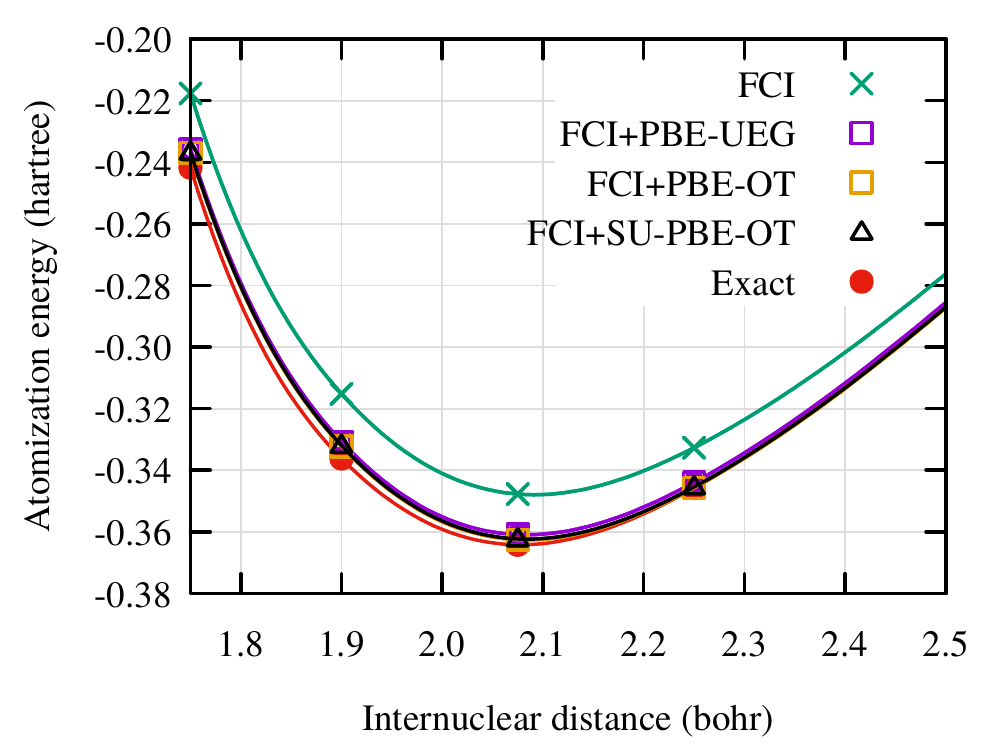}
     }
        \caption{
        Potential energy curves of the \ce{N2} molecule calculated with exFCI and basis-set corrected exFCI using the aug-cc-pVDZ (top) and aug-cc-pVTZ (bottom) basis sets. The estimated exact energies are based on a fit of experimental data and obtained from Ref.~\onlinecite{LieCle-JCP-74a}.
        \label{fig:N2}}
\end{figure*}
%%%%%%%%%%%%%%%%%%%%%%%%%%%%%%%%%%%%%%%%%%%%%%%%%%%%%%%%%%%%%%%%%%%%%%%%%%%%%%%%%%%%%%%%%%%%%%%%%%%%%%%%%%%%%%%%%%%%%%%%%%%%%%%%%

%%%%%%%%%%%%%%%%%%%%%%%%%%%%%%%%%%%%%%%%%%%%%%%%%%%%%%%%%%%%%%%%%%%%%%%%%%%%%%%%%%%%%%%%%%%%%%%%%%%%%%%%%%%%%%%%%%%%%%%%%%%%%%%%%
\begin{figure*}
 	\subfigure[aug-cc-pVDZ]{
       \includegraphics[width=0.45\linewidth] {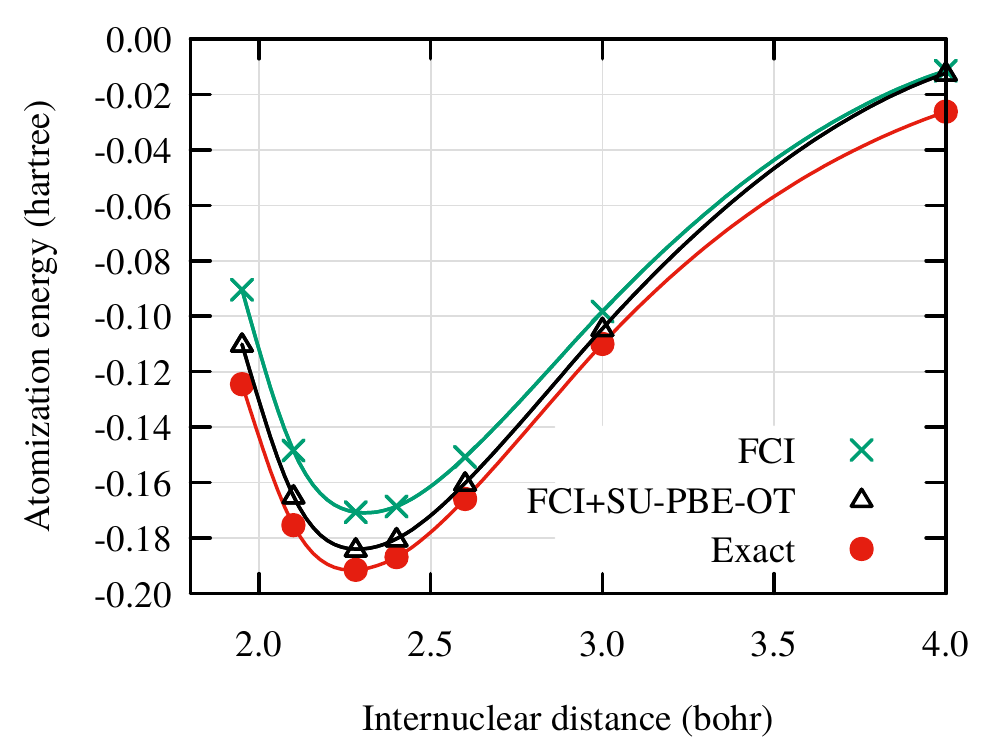}
        \includegraphics[width=0.45\linewidth]{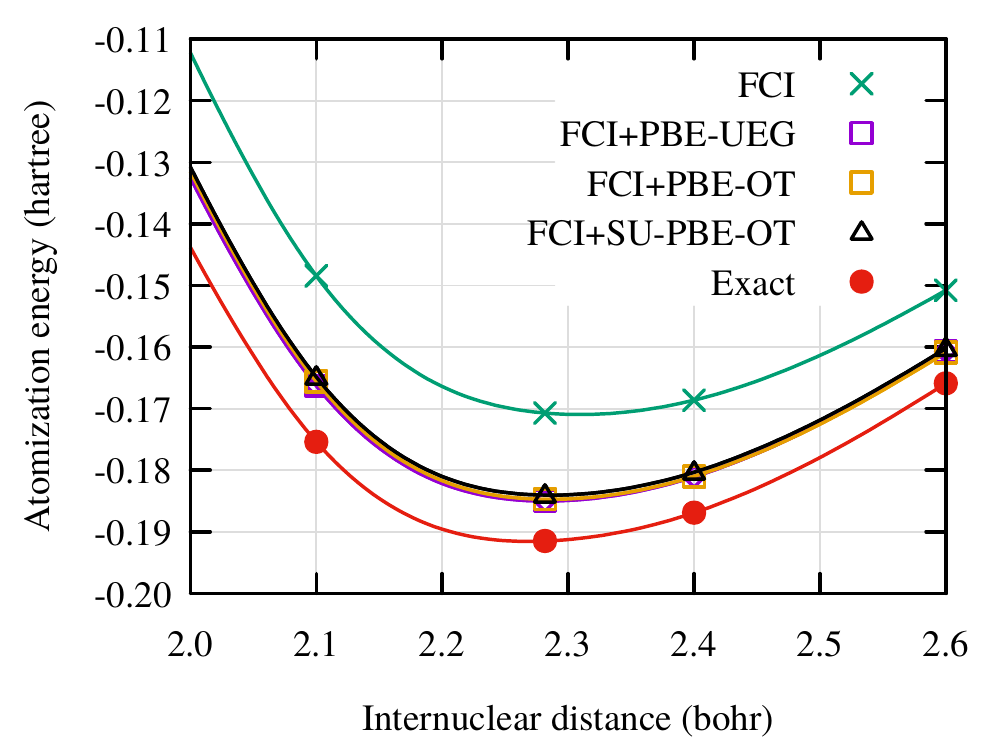}
	}
 	\subfigure[aug-cc-pVTZ]{
       \includegraphics[width=0.45\linewidth]  {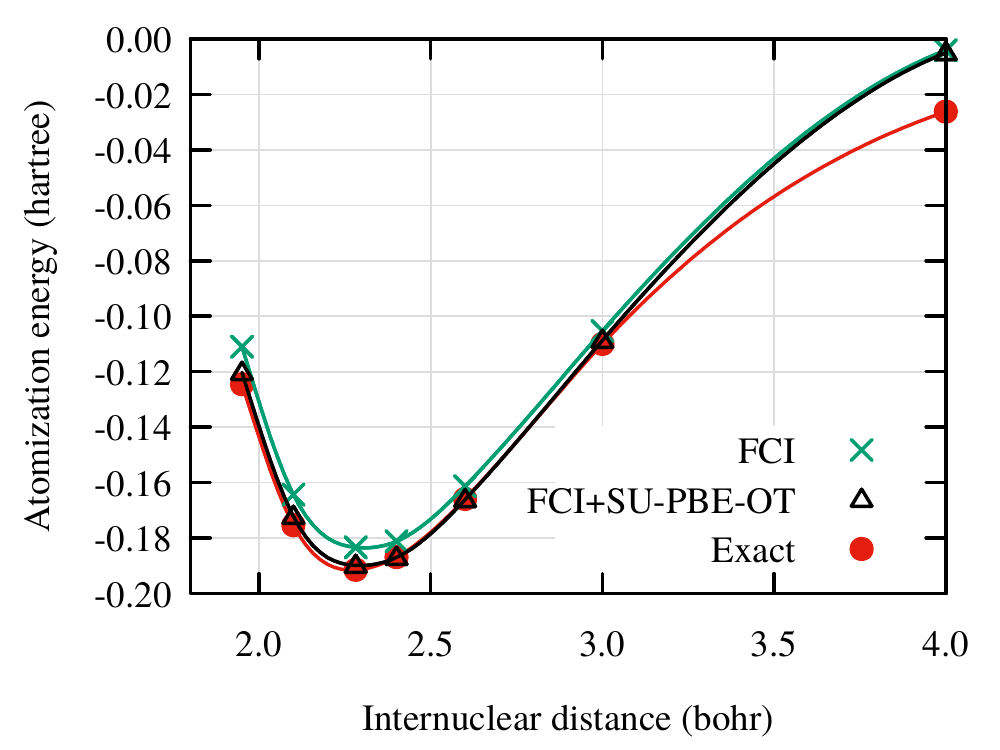}
        \includegraphics[width=0.45\linewidth] {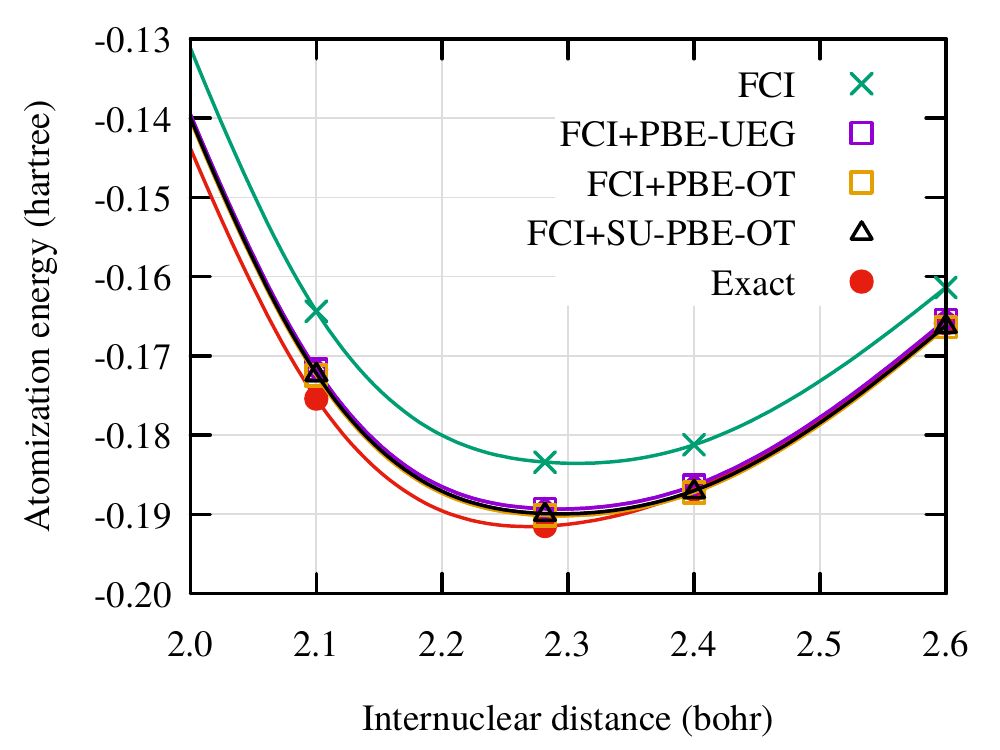}
	}
        \caption{
        Potential energy curves of the \ce{O2} molecule calculated with exFCI and basis-set corrected exFCI using the aug-cc-pVDZ (top) and aug-cc-pVTZ (bottom) basis sets.
The estimated exact energies are based on a fit of experimental data and obtained from Ref.~\onlinecite{LieCle-JCP-74a}.
        \label{fig:O2}}
\end{figure*}
%%%%%%%%%%%%%%%%%%%%%%%%%%%%%%%%%%%%%%%%%%%%%%%%%%%%%%%%%%%%%%%%%%%%%%%%%%%%%%%%%%%%%%%%%%%%%%%%%%%%%%%%%%%%%%%%%%%%%%%%%%%%%%%%%

%%%%%%%%%%%%%%%%%%%%%%%%%%%%%%%%%%%%%%%%%%%%%%%%%%%%%%%%%%%%%%%%%%%%%%%%%%%%%%%%%%%%%%%%%%%%%%%%%%%%%%%%%%%%%%%%%%%%%%%%%%%%%%%%%
\begin{figure*}
 	\subfigure[aug-cc-pVDZ]{
        \includegraphics[width=0.45\linewidth]{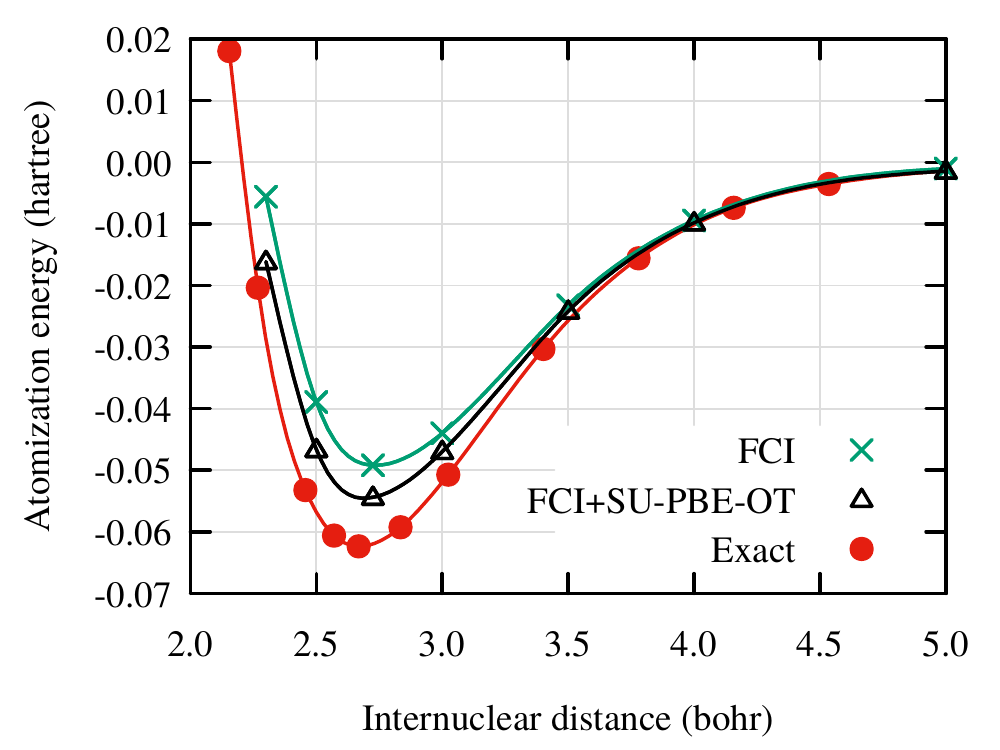}
        \includegraphics[width=0.45\linewidth]{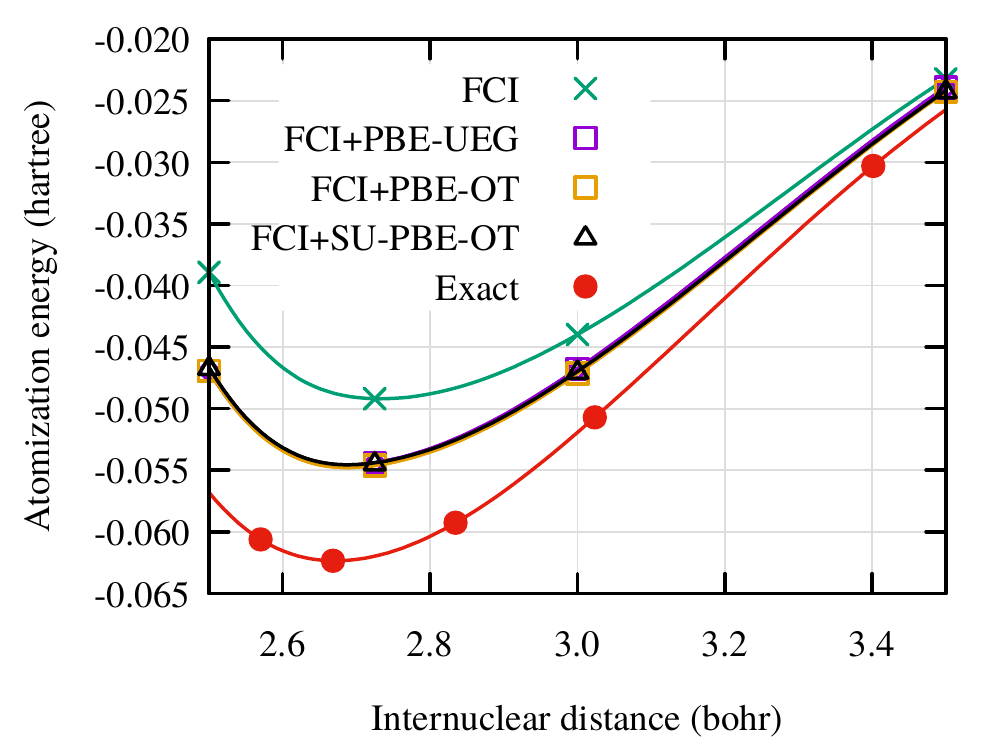}
	}
	\subfigure[aug-cc-pVTZ]{
        \includegraphics[width=0.45\linewidth]{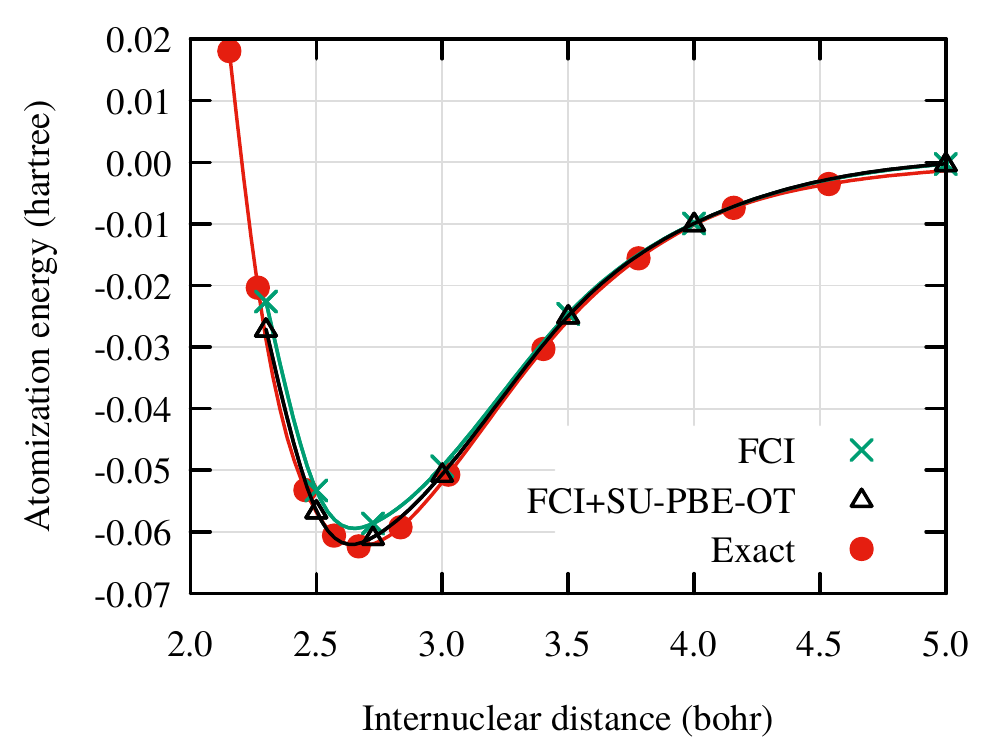}
        \includegraphics[width=0.45\linewidth]{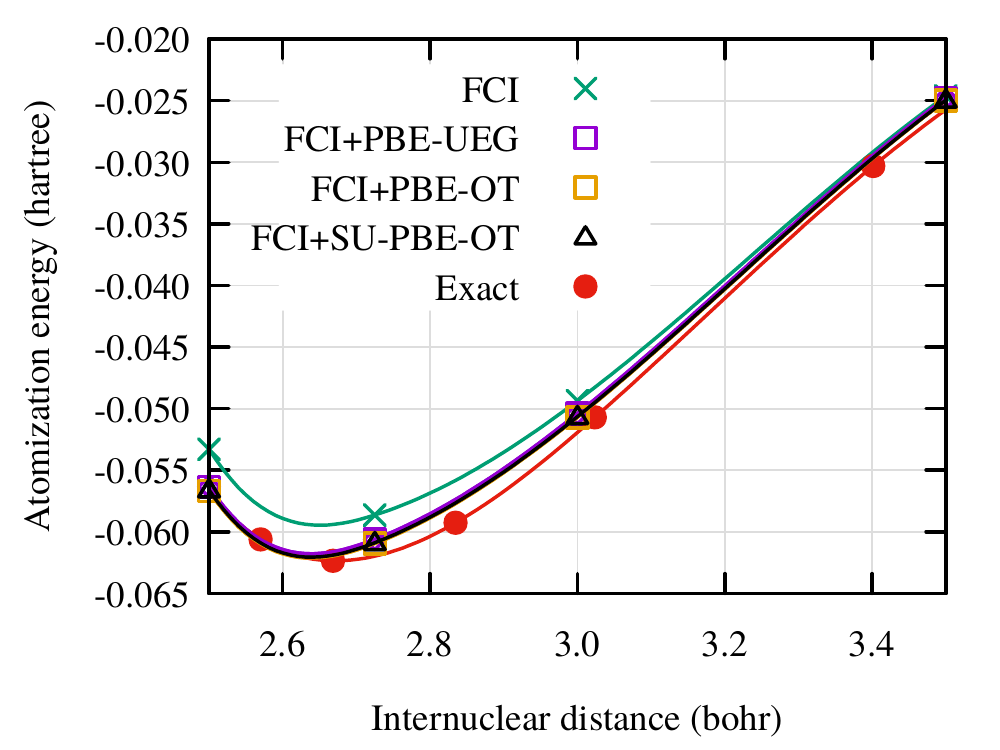}
	}
        \caption{
        Potential energy curves of the \ce{F2} molecule calculated with exFCI and basis-set corrected exFCI using the aug-cc-pVDZ (top) and aug-cc-pVTZ (bottom) basis sets.
The estimated exact energies are based on a fit of experimental data and obtained from Ref.~\onlinecite{LieCle-JCP-74a}.
        \label{fig:F2}}
\end{figure*}
%%%%%%%%%%%%%%%%%%%%%%%%%%%%%%%%%%%%%%%%%%%%%%%%%%%%%%%%%%%%%%%%%%%%%%%%%%%%%%%%%%%%%%%%%%%%%%%%%%%%%%%%%%%%%%%%%%%%%%%%%%%%%%%%%

We report in Figs.~\ref{fig:N2}, \ref{fig:O2}, and \ref{fig:F2} the potential energy curves of \ce{N2}, \ce{O2}, and \ce{F2} computed at various approximation levels using the aug-cc-pVDZ and aug-cc-pVTZ basis sets. The atomization energies for each level of theory with different basis sets are reported in Table \ref{tab:d0}.

Just as in \ce{H10}, the accuracy of the atomization energies is globally improved by adding the basis-set correction and it is remarkable that $\pbeontXi$ and $\pbeontns$ provide again very similar results. The latter observation confirms that the dependence on the on-top pair density allows one to remove the dependence of any kind of spin polarization for a quite wide range of covalent bonds and also for an open-shell system like \ce{O2}. More quantitatively, an error below 1.0 mHa compared to the estimated exact valence-only atomization energy is found for \ce{N2}, \ce{O2}, and \ce{F2} with the aug-cc-pVTZ basis set using the $\pbeontns$ functional, whereas such a feat is far from being reached within the same basis set at the near-FCI level. In the case of \ce{F2} it is clear that the addition of diffuse functions in the double- and triple-$\zeta$ basis sets strongly improves the accuracy of the results, which could have been anticipated due to the strong breathing-orbital effect induced by the ionic valence-bond forms in this molecule. \cite{HibHumByrLen-JCP-94} 
It should be also noticed that when reaching the aug-cc-pVQZ basis set for \ce{N2}, the accuracy of the atomization energy slightly deteriorates for the $\pbeontXi$ and $\pbeontns$ functionals, but it remains nevertheless more accurate than the estimated FCI atomization energy and very close to chemical accuracy. 

The overestimation of the basis-set-corrected atomization energy observed for \ce{N2} in large basis sets reveals an unbalanced treatment between the molecule and the atom in favor of the molecular system. Since the integral over $\br{}$ of the on-top pair density $\n{2}{}(\br{})$ is proportional to the short-range correlation energy in the large-$\mu$ limit \cite{PazMorGorBac-PRB-06,FerGinTou-JCP-18} [see Eq.~\eqref{eq:lim_mularge}], the accuracy of a given approximation of the exact on-top pair density will have a direct influence on the accuracy of the related basis-set correction energy $\bar{E}^\Bas$. To quantify the quality of different flavors of on-top pair densities for a given system and a given basis set $\basis$, we define the system-averaged CASSCF on-top pair density and extrapolated on-top pair density
\begin{subequations}
\begin{gather}
 \ontopcas = \int \text{d}\br{}\, n_{2,\text{CASSCF}}(\br{}),
 \label{eq:ontopcas}
\\
 \ontopextrap = \int \text{d}\br{}\, \mathring{n}_{2,\text{CASSCF}}(\br{}),
 \label{eq:ontopextrap}
\end{gather}
\end{subequations}
where $\mathring{n}_{2,\text{CASSCF}}(\br{})=\ntwoextrap(n_{2,\text{CASSCF}}(\br{}),\murcas)$ [see Eq.~\eqref{eq:def_n2extrap}] and $\murcas$ is the local range-separation function calculated with the CASSCF wave function, and similarly the system-averaged CIPSI on-top pair density and extrapolated on-top pair density
\begin{subequations}
\begin{align}
 \ontopcipsi & = \int \text{d}\br{}\,n_{2,\text{CIPSI}}(\br{}), 
 \label{eq:ontopcipsi}
\\
 \ontopextrapcipsi & = \int \text{d}\br{}\, \mathring{n}_{2,\text{CIPSI}}(\br{}), 
 \label{eq:ontopextrapcipsi}
\end{align}
\end{subequations}
where $\mathring{n}_{2,\text{CISPI}}(\br{})=\ntwoextrap(n_{2,\text{CIPSI}}(\br{}),\murcipsi)$ and $\murcipsi$ is the local range-separation function calculated with the CIPSI wave function. We also define the system-averaged range-separation parameters
\begin{subequations}
\begin{align}
 \muaverage & = \frac{1}{N}\int \text{d}\br{}\,n_{\text{CASSCF}}(\br{}) \,\, \murcas,
 \label{eq:muaverage}
\\
 \muaveragecipsi & = \frac{1}{N}\int \text{d}\br{}\,n_{\text{CIPSI}}(\br{}) \,\, \murcipsi, 
 \label{eq:muaveragecipsi}
\end{align}
\end{subequations}
where $n_{\text{CASSCF}}(\br{})$ and $n_{\text{CIPSI}}(\br{})$ are the CASSCF and CIPSI densities, respectively. All the CIPSI quantities have been calculated with the largest variational wave function computed in the CIPSI calculation with a given basis, which contains here at least $10^7$ Slater determinants. In particular, $\murcipsi$ has been calculated from Eqs.~\eqref{eq:def_mur_val}--\eqref{eq:twordm_val} with the opposite-spin two-body density matrix $\Gam{pq}{rs}$ of the largest variational CIPSI wave function for a given basis. All quantities in Eqs.~\eqref{eq:ontopcas}--\eqref{eq:muaverage} were computed excluding all contributions from the 1s orbitals, \ie, they are ``valence-only'' quantities. 

\begin{table*}
\caption{System-averaged on-top pair density $\langle n_2 \rangle$, extrapolated on-top pair density $\langle \mathring{n}_{2} \rangle$, and range-separation parameter $\langle \mu \rangle$ (all in atomic units) calculated with full-valence CASSCF and CIPSI wave functions (see text for details) for \ce{N2} and \ce{N} in the aug-cc-pVXZ basis sets (X $=$ D, T, and Q). All quantities were computed within the frozen-core approximation, \ie, excluding all contributions from the 1s orbitals.}
\begin{ruledtabular}
\begin{tabular}{lrccccccc}
%\begin{tabular}{lrccccccc}

   System 	&  \tabc{Basis set} &\tabc{$\ontopcas$}& \tabc{$\ontopextrap$}& \tabc{$\ontopcipsi$} & \tabc{$\ontopextrapcipsi$}& \tabc{$\muaverage$}  & \tabc{$\muaveragecipsi$}       \\
\hline                                                                                                                                                                  
\ce{N2}         &  aug-cc-pVDZ      &  1.17542         &   0.65966            &   1.02792            &   0.58228                 &   0.946              &   0.962                   \\
                &  aug-cc-pVTZ      &  1.18324         &   0.77012            &   0.92276            &   0.61074                 &   1.328              &   1.364                   \\
                &  aug-cc-pVQZ      &  1.18484         &   0.84012            &   0.83866            &   0.59982                 &   1.706              &   1.746                   \\[0.1cm]
                                                                                                                                                                                   
\ce{N}          &  aug-cc-pVDZ      &  0.34464         &   0.19622            &   0.25484            &   0.14686                 &   0.910              &   0.922                   \\
                &  aug-cc-pVTZ      &  0.34604         &   0.22630            &   0.22344            &   0.14828                 &   1.263              &   1.299                   \\
                &  aug-cc-pVQZ      &  0.34614         &   0.24666            &   0.21224            &   0.15164                 &   1.601              &   1.653                   \\
\end{tabular}
\end{ruledtabular}
\label{tab:d1}
\end{table*}

We report in Table \ref{tab:d1} these quantities for \ce{N2} and \ce{N} for various basis sets. One notices that the system-averaged on-top pair density at the CIPSI level $\ontopcipsi$ is systematically lower than its CASSCF analogue $\ontopcas$, which is expected since short-range correlation, \ie, digging the correlation hole in a given basis set at near FCI level, is missing from the valence CASSCF wave function. 
Also, $\ontopcipsi$ decreases in a monotonous way as the size of the basis set increases, leading to roughly a $20\%$ decrease from the aug-cc-pVDZ to the aug-cc-pVQZ basis sets, whereas $\ontopcas$ is almost constant with respect to the basis set. Regarding the extrapolated on-top pair densities, $\ontopextrap$ and $\ontopextrapcipsi$, it is interesting to notice that they are substantially lower than their non-extrapolated counterparts, $\ontopcas$ and $\ontopcipsi$. Nevertheless, the behaviors of $\ontopextrap$ and $\ontopextrapcipsi$ are qualitatively different: $\ontopextrap$ clearly increases when enlarging the basis set whereas $\ontopextrapcipsi$ remains almost constant. More precisely, in the case of \ce{N2}, the value of $\ontopextrap$ increases by about 30$\%$ from the aug-cc-pVDZ to the aug-cc-pVQZ basis sets, whereas the value of $\ontopextrapcipsi$ only fluctuates within 5$\%$ for the same basis sets. The behavior of $\ontopextrap$ can be understood by noticing that i) the value of $\murcas$ globally increases when enlarging the basis set (as evidenced by $\muaverage$), and ii) $\lim_{\mu \rightarrow \infty} \ntwoextrap(n_2,\mu) = n_2$ [see Eq.~\eqref{eq:def_n2extrap}]. Therefore, in the CBS limit, $\murcas \rightarrow \infty$ and one obtains 
\begin{equation}
 \lim_{\basis \rightarrow \text{CBS}} \ontopextrap =  \lim_{\basis \rightarrow \text{CBS}} \ontopcas,
\end{equation}
\ie, $\ontopextrap$ must increase with the size of the basis set $\basis$ to eventually converge to $\lim_{\basis \rightarrow \text{CBS}} \ontopcas$, the latter limit being essentially reached with the present basis sets.
On the other hand, the stability of $\ontopextrapcipsi$ with respect to the basis set is quite remarkable and must come from the fact that i) $\ontopcipsi$ is a good approximation to the corresponding FCI value within the considered basis sets, and ii) the extrapolation formula in Eq.~\eqref{eq:def_n2extrap} together with the choice of $\murcipsi$ are quantitatively correct. Therefore, we expect the calculated values of $\ontopextrapcipsi$ to be nearly converged with respect to the basis set, and we will take the value of $\ontopextrapcipsi$ in the aug-cc-pVQZ basis set as an estimate of the exact system-averaged on-top pair density.

For the present work, it is important to keep in mind that $\ontopextrap$ directly determines the basis-set correction in the large-$\mu$ limit. More precisely, the correlation energy contribution associated with the basis-set correction is (in absolute value) an increasing function of $\ontopextrap$. Therefore, the error on $\ontopextrap$ with respect to the estimated exact system-averaged on-top pair density provides an indication of the error made by the basis-set correction for a given system and basis set. With the aug-cc-pVQZ basis set, we have $\ontopextrap - \ontopextrapcipsi = 0.240$ for the \ce{N2} molecule, while $2(\ontopextrap - \ontopextrapcipsi) = 0.190$ for two isolated \ce{N} atoms. We can then conclude that the overestimation of the system-averaged on-top pair density, and therefore of the basis-set correction, is more important for the \ce{N2} molecule at equilibrium distance than for the isolated \ce{N} atoms. This probably explains the observed overestimation of the atomization energy. To confirm this statement, we computed the basis-set correction for both the \ce{N2} molecule at equilibrium distance and the isolated atoms using $\murcipsi$ and $\mathring{n}_{2,\text{CIPSI}}(\br{})$ with the aug-cc-pVTZ and aug-cc-pVQZ basis sets. We obtained the following values for the atomization energies: $362.12$ mH with aug-cc-pVTZ and $362.15$ mH with aug-cc-pVQZ, which are indeed more accurate values than those obtained using $\murcas$ and $\mathring{n}_{2,\text{CASSCF}}(\br{})$.

Finally, regarding now the performance of the basis-set correction along the whole potential energy curves reported in Figs.~\ref{fig:N2}, \ref{fig:O2}, and \ref{fig:F2}, it is interesting to notice that it fails to provide a noticeable improvement far from the equilibrium geometry. Acknowledging that the weak-correlation effects in these regions are dominated by dispersion interactions which are long-range effects, the failure of the present approximations for the complementary functional can be understood easily. Indeed, the whole scheme designed here is based on the physics of correlation near the electron-electron coalescence point: the local range-separation function $\mu(\br{})$ is based on the value of the effective electron-electron interaction at coalescence and the ECMD functionals are suited for short-range correlation effects. Therefore, the failure of the present basis-set correction to describe dispersion interactions is theoretically expected.
We hope to report further on this in the near future.

\section{Conclusion}
\label{sec:conclusion}

In the present paper we have extended the recently proposed DFT-based basis-set correction to strongly correlated systems. We have applied the method to the \ce{H10}, \ce{N2}, \ce{O2}, and \ce{F2} molecules up to the dissociation limit at near-FCI level in increasingly large basis sets, and investigated how the basis-set correction affects the convergence toward the CBS limit of the potential energy curves of these molecular systems. 

The density-based basis-set correction relies on three aspects: i) the definition of an effective non-divergent electron-electron interaction obtained from the expectation value over a wave function $\psibasis$ of the Coulomb electron-electron interaction projected into an incomplete basis set $\basis$; ii) the fit of this effective interaction with the long-range interaction used in RSDFT; and iii) the use of a short-range, complementary functional borrowed from RSDFT. In the present paper, we investigated i) and iii) in the context of strong correlation and focused on potential energy curves  and atomization energies. More precisely, we proposed a new scheme to design functionals fulfilling spin-multiplet degeneracy and size consistency. To fulfill such requirements we proposed to use CASSCF wave functions leading to size-consistent energies, and we developed functionals using only $S_z$-independent density-like quantities. 

The development of new $S_z$-independent and size-consistent functionals has lead us to investigate the role of two related quantities: the spin polarization and the on-top pair density. One important result of the present study is that by using functionals \textit{explicitly} depending on the on-top pair density, one can eschew its spin-polarization dependence without loss of accuracy. This avoids the commonly used effective spin polarization originally proposed in Ref.~\onlinecite{BecSavSto-TCA-95} which has the disadvantage of possibly becoming complex-valued in the multideterminant case. From a more fundamental aspect, this confirms that, in a DFT framework, the spin polarization mimics the role of the on-top pair density. 
Consequently, we believe that one could potentially develop new families of density-functional approximations where the spin polarization is abandoned and replaced by the on-top pair density.

Regarding the results of the present approach, the basis-set correction systematically improves the near-FCI calculations in a given basis set. More quantitatively, it is shown that with only triple-$\zeta$ quality basis sets chemically accurate atomization energies are obtained for all systems whereas the uncorrected near-FCI results are far from this accuracy within the same basis set. 

Also, it is shown that the basis-set correction gives substantial differential contribution to potential energy curves close to the equilibrium geometries, but at long internuclear distances it cannot recover the dispersion interaction energy missing because of the basis-set incompleteness. This behavior is actually expected as dispersion interactions are of long-range nature and the present approach is designed to recover only short-range correlation effects.

\appendix

\section{Size consistency of the basis-set correction}
\label{app:sizeconsistency}

\subsection{Sufficient condition for size consistency}

The basis-set correction is expressed as an integral in real space
\begin{multline}
 \label{eq:def_ecmdpbebasisAnnex}
 \efuncdenpbe{\argebasis} = \\ \int  \text{d}\br{} \,\denr   \ecmd(\argrebasis),
\end{multline}
where all the local quantities $\argrebasis$ are obtained from the same wave function $\Psi$. In the limit of two non-overlapping and non-interacting dissociated fragments $\text{A}+\text{B}$, this integral can be rewritten as the sum of the integral over the region $\Omega_\text{A}$ and the integral over the region $\Omega_\text{B}$
\begin{multline}
 \label{eq:def_ecmdpbebasisAB}
 \efuncdenpbeAB{\argebasis} = 
 \\ 
 \int_{\Omega_\text{A}}  \text{d}\br{} \,\denr   \ecmd(\argrebasis) 
 \\ 
 + \int_{\Omega_\text{B}}  \text{d}\br{} \,\denr   \ecmd(\argrebasis).
\end{multline}
Therefore, a sufficient condition to obtain size consistency is that all the local quantities $\argrebasis$ are \textit{intensive}, \ie, they \textit{locally} coincide in the supersystem $\text{A}+\text{B}$ and in each isolated fragment $\text{X}=\text{A}$ or $\text{B}$. Hence, we must have, for $\br{} \in \Omega_\text{X}$, 
\begin{subequations}
\begin{gather}
n_\text{A+B}(\br{}) = n_\text{X}(\br{}),
\label{nAB}
\\
\zeta_\text{A+B}(\br{}) = \zeta_\text{X}(\br{}),
\label{zAB}
\\
s_\text{A+B}(\br{}) = s_\text{X}(\br{}),
\label{sAB}
\\
n_{2,\text{A+B}}(\br{}) = n_{2,\text{X}}(\br{}),
\label{n2AB}
\\
\mu_{\text{A+B}}(\br{}) = \mu_{\text{X}}(\br{}),
\label{muAB}
\end{gather}
\end{subequations}
where the left-hand-side quantities are for the supersystem and the right-hand-side quantities for an isolated fragment. Such conditions can be difficult to fulfill in the presence of degeneracies since the system X can be in a different mixed state (\ie, ensemble) in the supersystem $\text{A}+\text{B}$ and in the isolated fragment. \cite{Sav-CP-09} Here, we will consider the simple case where the wave function of the supersystem is multiplicatively separable, \ie,
\begin{equation}
 \ket*{\wf{\text{A}+\text{B}}{}} = \ket*{\wf{\text{A}}{}} \otimes \ket*{\wf{\text{B}}{}}, 
\label{PsiAB}
\end{equation}
where $\otimes$ is the antisymmetric tensor product. In this case, it is easy to shown that Eqs.~(\ref{nAB})-(\ref{sAB}) are valid, as well known, and it remains to show that Eqs.~(\ref{n2AB}) and~(\ref{muAB}) are also valid. Before showing this, we note that even though we do not explicitly consider the case of degeneracies, the lack of size consistency which could arise from spin-multiplet degeneracies can be avoided by the same strategy used for imposing the energy independence on $S_z$, \ie, by using the effective spin polarization $\tilde{\zeta}(n(\br{}),n_{2}(\br{}))$ or a zero spin polarization $\zeta(\br{}) = 0$. Moreover, for the systems treated in this work, the lack of size consistency which could arise from spatial degeneracies (coming from atomic $p$ states) can also be avoided by selecting the same state in the supersystem and in the isolated fragment. For example, for the \ce{F2} molecule, the CASSCF wave function dissociates into the atomic configuration $\text{p}_\text{x}^2 \text{p}_\text{y}^2 \text{p}_\text{z}^1$ for each fragment, and we thus choose the same configuration for the calculation of the isolated atom. The same argument applies to the \ce{N2} and \ce{O2} molecules. For other systems, it may not be always possible to do so.

\subsection{Intensivity of the on-top pair density and the local range-separation function}

The on-top pair density can be written in an orthonormal spatial orbital basis set $\{\SO{p}{}\}$ as
\begin{equation}
\label{eq:def_n2}
 n_{2{}}(\br{})  = \sum_{pqrs \in \Bas} \SO{p}{} \SO{q}{} \Gam{pq}{rs}  \SO{r}{} \SO{s}{}, 
\end{equation}
with $\Gam{pq}{rs} = 2 \mel*{\wf{}{}}{ \aic{r_\downarrow}\aic{s_\uparrow}\ai{q_\uparrow}\ai{p_\downarrow}}{\wf{}{}}$. As the summations run over all orbitals in the basis set $\Bas$, $n_{2{}}(\br{})$ is invariant to orbital rotations and can thus be expressed in terms of localized orbitals. For two non-overlapping fragments $\text{A}+\text{B}$, the basis set can then be partitioned into orbitals localized on the fragment A and orbitals localized on B, \ie, $\Bas=\Bas_\text{A}\cup \Bas_\text{B}$. Therefore, we see that the on-top pair density of the supersystem $\text{A}+\text{B}$ is additively separable
\begin{equation}
 n_{2,\text{A}+\text{B}}(\br{})  = n_{2,\text{A}}(\br{}) + n_{2,\text{B}}(\br{}),
\end{equation}
where $n_{2,\text{X}}(\br{})$ is the on-top pair density of the fragment X
\begin{equation}
 n_{2,\text{X}}(\br{})  = \sum_{pqrs \in \Bas_\text{X}} \SO{p}{} \SO{q}{} \Gam{pq}{rs}  \SO{r}{} \SO{s}{}, 
\end{equation}
in which the elements $\Gam{pq}{rs}$ with orbital indices restricted to the fragment X are $\Gam{pq}{rs} = 2 \mel*{\wf{\text{A}+\text{B}}{}}{ \aic{r_\downarrow}\aic{s_\uparrow}\ai{q_\uparrow}\ai{p_\downarrow}}{\wf{\text{A}+\text{B}}{}} = 2 \mel*{\wf{\text{X}}{}}{ \aic{r_\downarrow}\aic{s_\uparrow}\ai{q_\uparrow}\ai{p_\downarrow}}{\wf{\text{X}}{}}$, owing to the multiplicative structure of the wave function [see Eq.~\eqref{PsiAB}]. This shows that the on-top pair density is a local intensive quantity.

The local range-separation function is defined as, for $n_{2}(\br{}) \not=0$,
\begin{equation}
 \label{eq:def_murAnnex}
 \mur = \frac{\sqrt{\pi}}{2} \frac{f(\bfr{},\bfr{})}{n_{2}(\br{})},
\end{equation}
where 
\begin{equation}
 \label{eq:def_f}
    f(\bfr{},\bfr{}) = \sum_{pqrstu\in \Bas} \SO{p}{ } \SO{q}{ } \V{pq}{rs} \Gam{rs}{tu} \SO{t}{ } \SO{u}{ }.
\end{equation}
Again, $f(\bfr{},\bfr{})$ is invariant to orbital rotations and can be expressed in terms of orbitals localized on the fragments A and B. In the limit of infinitely separated fragments, the Coulomb interaction vanishes between A and B and therefore any two-electron integral $\V{pq}{rs}$ involving orbitals on both A and B vanishes. We thus see that the quantity $f(\bfr{},\bfr{})$ of the supersystem $\text{A}+\text{B}$ is additively separable
\begin{equation}
 \label{eq:def_fa+b}
    f_{\text{A}+\text{B}}(\bfr{},\bfr{}) =  f_{\text{A}}(\bfr{},\bfr{})  +  f_{\text{B}}(\bfr{},\bfr{}),
\end{equation}
with 
\begin{equation}
 \label{eq:def_fX}
    f_\text{X}(\bfr{},\bfr{}) = \sum_{pqrstu\in \Bas_\text{X}} \SO{p}{ } \SO{q}{ } \V{pq}{rs} \Gam{rs}{tu} \SO{t}{ } \SO{u}{ }.
\end{equation}
So, $f(\bfr{},\bfr{})$ is a local intensive quantity.
As a consequence, the local range-separation function of the supersystem $\text{A}+\text{B}$ is
\begin{equation}
 \label{eq:def_murAB}
 \mu_{\text{A}+\text{B}}(\bfr{}) = \frac{\sqrt{\pi}}{2} \frac{f_{\text{A}}(\bfr{},\bfr{})  +  f_{\text{B}}(\bfr{},\bfr{})}{n_{2,\text{A}}(\br{}) + n_{2,\text{B}}(\br{})},
\end{equation}
which implies
\begin{equation}
 \label{eq:def_murABsum}
 \mu_{\text{A}+\text{B}}(\bfr{}) = \mu_{\text{X}}(\bfr{}) \;\; \text{if} \;\; \bfr{} \in \Omega_\text{X},
\end{equation}
where $\mu_{\text{X}}(\bfr{}) = (\sqrt{\pi}/2) f_{\text{X}}(\bfr{},\bfr{})/n_{2,\text{X}}(\br{})$.
The local range-separation function is thus a local intensive quantity.

We can therefore conclude that, if the wave function of the supersystem  $\text{A}+\text{B}$ is multiplicative separable, all local quantities used in the basis-set correction functional are intensive and therefore the basis-set correction is size consistent.

\section{Computational cost of the basis-set correction for a CASSCF wave function}
\label{app:computational}

The computational cost of the basis-set correction is determined by the calculation of the on-top pair density $n_{2}(\br{})$ and the local range-separation function $\mur$ on the real-space grid. For a general multideterminant wave function, the computational cost is of order $O(N_\text{grid}N_{\Bas}^4)$ where $N_\text{grid}$ is the number of grid points and $N_{\Bas}$ is the number of basis functions.\cite{LooPraSceTouGin-JCPL-19} For a CASSCF wave function, a significant reduction of the scaling of the computational cost can be achieved.

\subsection{Computation of the on-top pair density}

For a CASSCF wave function $\Psi$, the occupied orbitals can be partitioned into a set of active orbitals $\mathcal{A}$ and a set of inactive (doubly occupied) orbitals $\mathcal{I}$. The CASSCF on-top pair density can then be written as 
\begin{equation}
 \label{def_n2_good}
 n_{2}(\br{})  = n_{2,\mathcal{A}}(\br{}) + n_{\mathcal{A}}(\br{}) n_{\mathcal{I}}(\br{}) + \frac{n_{\mathcal{I}}(\br{})^2}{2},
\end{equation}
where
\begin{subequations}
\begin{align}
 \label{def_n2_act}
 n_{2,\mathcal{A}}(\br{})  & = \sum_{pqrs \in \mathcal{A}} \SO{p}{} \SO{q}{} \Gam{pq}{rs}  \SO{r}{} \SO{s}{}, 
	\\
 n_{\mathcal{A}}(\br{}) &  = \sum_{pq \in\mathcal{A}} \phi_p (\br{})  \phi_q (\br{}) 
 \mel*{\wf{}{}}{ \aic{p_\uparrow}\ai{q_\uparrow} + \aic{p_\downarrow}\ai{q_\downarrow} }{\wf{}{}},
 \\
 n_{\mathcal{I}}(\br{}) & = 2 \sum_{p \in \mathcal{I}} \phi_p (\br{})^2
\end{align}
\end{subequations}
are the purely active part of the on-top pair density, the active part of the density, and the inactive part of the density, respectively. 
The leading computational cost is the evaluation of $n_{2,\mathcal{A}}(\br{})$ on the grid which, according to Eq.~\eqref{def_n2_act}, scales as $O(N_\text{grid} N_\mathcal{A}^4)$ where $N_{\mathcal{A}}$ is the number of active orbitals which is much smaller than the number of basis functions $N_{\Bas}$.

\subsection{Computation of the local range-separation function}

In addition to the on-top pair density, the computation of $\mur$ needs the computation of $f(\bfr{},\bfr{})$ [see Eq.~\eqref{eq:def_f}] at each grid point. It can be factorized as 
\begin{equation}
 \label{eq:f_good}
 f(\bfr{},\bfr{})  =  \sum_{rs \in \Bas} V^{rs}(\bfr{}) \, \Gamma_{rs}(\bfr{}),
\end{equation}
where 
\begin{subequations}
\begin{align}
 V^{rs}(\bfr{}) & = \sum_{pq \in \Bas} V_{pq}^{rs} \phi_p(\br{}) \phi_q(\br{}),
 \\
 \Gamma_{rs}(\bfr{}) & = \sum_{pq \in \Bas} \Gam{rs}{pq} \phi_p(\br{}) \phi_q(\br{}) .
\end{align}
\end{subequations}
For a general multideterminant wave function, the computational cost of $f(\bfr{},\bfr{})$ thus scales as $O(N_\text{grid}N_{\Bas}^4)$.

In the case of a CASSCF wave function, $\Gam{rs}{pq}$ vanishes if one index $p,q,r,s$ does not belong to the set of inactive or active occupied orbitals $\mathcal{I}\cup \mathcal{A}$. Therefore, at a given grid point, the number of non-zero elements $\Gamma_{rs}(\bfr{})$ is only at most $(N_{\mathcal{I}}+N_{\mathcal{A}})^2$, which is usually much smaller than $N_{\Bas}^2$. As a consequence, one can also restrict the sum in the calculation of
\begin{equation}
 f(\bfr{},\bfr{}) = \sum_{rs \in \mathcal{I}\cup\mathcal{A}} V^{rs}(\bfr{}) \, \Gamma_{rs}(\bfr{}). 
\end{equation}
The overall computational cost is dominated by that of $V^{rs}(\bfr{})$, which scales as $O(N_\text{grid}(N_{\mathcal{I}}+N_{\mathcal{A}})^2 N_{\Bas}^2)$, which is much smaller than $O(N_\text{grid}N_{\Bas}^4)$.

\end{document}